\documentclass[prx,twocolumn,longbibliography,superscriptaddress,preprintnumbers]{revtex4-2}
\usepackage[colorlinks,bookmarks=true,citecolor=blue,linkcolor=blue,urlcolor=blue]{hyperref}
\usepackage{dcolumn,graphicx,amsfonts,amsthm,bm,color,appendix,float,tabularx}
\usepackage{braket}
\usepackage[normalem]{ulem}
\usepackage[version=4]{mhchem}
\usepackage{siunitx}
\usepackage{amsmath}
\usepackage{amssymb}
\usepackage{CJKutf8}
\usepackage{multirow}

\usepackage[dvipsnames]{xcolor}
\definecolor{pal0}{rgb}{0.8941, 0.102 , 0.1098}
\definecolor{pal1}{rgb}{0.2157, 0.4941, 0.7216}
\definecolor{pal2}{rgb}{0.302 , 0.6863, 0.2902}
\definecolor{pal3}{rgb}{0.5961, 0.3059, 0.6392}
\definecolor{pal4}{rgb}{1.    , 0.498 , 0.    }

\DeclareMathOperator{\Tr}{Tr}

\newcommand{\n}[1]{\left| #1 \right|}
\newcommand{\setc}[2]{\left\{#1\; :\; #2 \right\}}
\newcommand{\st}[1]{\left\{#1\right\}}
\renewcommand{\v}[1]{\boldsymbol{#1}}

\newcommand{\bz}{\mathrm{BZ}}

\newcommand{\PRLsec}[1]{\emph{#1---}}

\newcommand{\RNG}{RMG} 

\begin{document}

\title{
Anomalous Hall Crystals in Rhombohedral Multilayer Graphene I: Interaction-Driven Chern Bands and  Fractional Quantum Hall States at Zero Magnetic Field
}

\author{Junkai Dong (\begin{CJK*}{UTF8}{bsmi}董焌\end{CJK*}\begin{CJK*}{UTF8}{gbsn}锴\end{CJK*})}
\thanks{These authors contributed equally.}
\affiliation{Department of Physics, Harvard University, Cambridge, MA 02138, USA}

\author{Taige Wang}
\thanks{These authors contributed equally.}
\affiliation{Department of Physics, University of California, Berkeley, CA 94720, USA}
\affiliation{Material Science Division, Lawrence Berkeley National Laboratory, Berkeley, CA 94720, USA}

\author{Tianle Wang}
\thanks{These authors contributed equally.}
\affiliation{Department of Physics, University of California, Berkeley, CA 94720, USA}
\affiliation{Material Science Division, Lawrence Berkeley National Laboratory, Berkeley, CA 94720, USA}

\author{Tomohiro Soejima (\begin{CJK*}{UTF8}{bsmi}副島智大\end{CJK*})}
\affiliation{Department of Physics, Harvard University, Cambridge, MA 02138, USA}

\author{Michael P. Zaletel}
\affiliation{Department of Physics, University of California, Berkeley, CA 94720, USA}
\affiliation{Material Science Division, Lawrence Berkeley National Laboratory, Berkeley, CA 94720, USA}

\author{Ashvin Vishwanath}
\affiliation{Department of Physics, Harvard University, Cambridge, MA 02138, USA}

\author{Daniel E. Parker}
\affiliation{Department of Physics, University of California, Berkeley, CA 94720, USA}

\begin{abstract}
Recent experiments on rhombohedral pentalayer graphene with a substrate-induced moir\'e potential have identified both Chern insulators and fractional Quantum Hall states at zero magnetic field. Surprisingly, these states are observed in strong displacement fields where the effects of the moir\'e lattice are weak, and seem to be readily accessed without fine-tuning. To address these experimental puzzles, we study a model of interacting electrons in this geometry. Within self-consistent Hartree-Fock (SCHF), we find an isolated Chern band with small bandwidth and good quantum geometry. Exact diagonalization and density-matrix renormalization group calculations both confirm the band hosts fractional quantum Hall states without a magnetic field. Remarkably, the Chern band is stable at a wide range of angles, at four through six rhombohedral layers, at varying rhombohedral hopping parameters, and --- most strikingly --- survives in SCHF when the moir\'e potential vanishes. In this limit, the state spontaneously breaks time-reversal and translation symmetry simultaneously, giving a topological crystalline state that we term the ``anomalous Hall crystal'' (AHC). We argue this is a general mechanism to create stable Chern bands in rhombohedral multilayer graphene, opening the door to studying the interplay between electronic topology, fractionalization, and spontaneous translation symmetry breaking.
\end{abstract}

\maketitle


\begin{figure}[t]
    \includegraphics[width=\linewidth]{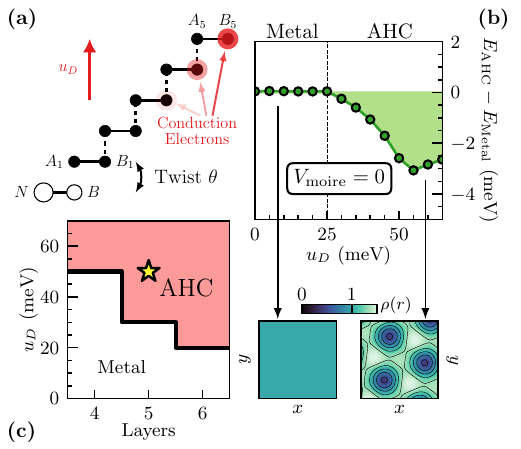}
    \caption{(a) Side view of rhombohedral pentalayer graphene aligned to an hBN substrate on the bottom with a twist angle $\theta$. The potential $u_D>0$ polarizes electrons away from the hBN, making them ``moir\'e-distant".
    (b) The energy difference per electron between a graphene translation symmetric metal and an anomalous Hall crystal (AHC) within SCHF as a function of $u_D$ in rhombohedral pentalayer graphene at \textit{zero} moir\'{e} potential $V_1 = 0$. The AHC is the ground state beyond $u_D=\SI{25}{\milli\electronvolt}$. Typical charge densities $\rho(r)$ (lower right) show spontaneous translation symmetry breaking in the AHC.
    (c) Phase diagram of rhombohedral multilayer graphene for $N_L=4 - 6$ layers with phase boundary from SCHF. The critical displacement field $u_D^*$ decreases with increasing $N_L$. Yellow star: estimated experimental parameters~\cite{lu2023fractional}.}
	\label{fig:intro}
\end{figure}

Tunable experimental platforms that host correlated topological states are fertile grounds for exotic phenomena. In recent years, a menagerie of candidates have arisen, including twisted graphene multilayers~\cite{xie2021fractional, cao2018correlated, Cory_Science19,
Efetov2020Screening, jaoui_quantum_2022, stepanov_competing_2021, cao2018unconventional, zondiner_cascade_2020, park_flavour_2021, cao2020_nematicity, Yankowitz2019, saito_independent_2020, wong_cascade_2020, oh_evidence_2021, choi2021_STM, yu_correlated_2022, chen2020monobi, polshyn2020electrical, polshyn2021topological, zhang2023local, cao2020doublebilayer, burg2019correlated, liu2020tunable, he2021symmetry, shen2020correlated, hao2021electric, park2021tunable, park2022robust, xia2023helical, uri2023quasicrystal, zhou2021superconductivity}, twisted transition metal dichalcogenide (TMD) bilayers~\cite{Li_2021_2,zhao2022realization,tao_valley-coherent_2022,foutty2023mapping,cai2023signatures, zeng2023thermodynamic, park2023observation, xu2023observation, anderson2023programming,wu_topological_2019,yu_giant_2020,zhai_theory_2020,tang_geometric_2021,devakul_magic_2021,zhang_spin-textured_2021,wang_staggered_2023,pan_band_2020,abouelkomsan_multiferroicity_2022,crepel_chiral_2023,li_electrically_2023,qiu_interaction-driven_2023,li_spontaneous_2021,crepel_anomalous_2023,morales-duran_pressure-enhanced_2023,wang_fractional_2023,reddy_fractional_2023,jia_moire_2023,yu_fractional_2023,wang_topological_2023,goldman_zero-field_2023,reddy_toward_2023,dong_composite_2023,morales-duran_magic_2023}, and rhombohedral multilayer graphene (RMG)~\cite{hallcrystal2022,Chen:2020aa,zhou2021half, lu2023fractional,han2023correlated,han2023large,Chen2022ferro,SpantonFCI, Zhou2022BLG,Zhang2023BLG,Holleis2023BLG,Chen2019SC, Lee2022TLG,Alexander20214LG,Liu20234LG,Han2023multiferroicity,sha2023observation,Senthil_NearlyFlatBand,Chittari2019flat,Repellin2020narrow,Zhang2019Hubbard}. Topological phenomena have been reported in RMG systems with $N_L =2-5$ layers, both with~\cite{Chen:2020aa,zhou2021half,Chen2022ferro} and without~\cite{hallcrystal2022,han2023correlated,sha2023observation,han2023large} moir\'e potentials due to an hBN substrate. A number of theoretical studies have examined this hBN moir\'e potential~\cite{moon2014electronic,JungAbInitio,jung2017moire, krisna2023moire, lee2016ballistic} and the resulting topological bands~\cite{Senthil_NearlyFlatBand, patri2023moire, Chittari2019flat,Repellin2020narrow,Zhang2019Hubbard, Cecile_PRL20}. A recent experiment~\cite{lu2023fractional} reported that rhombohedral pentalayer graphene aligned to an hBN substrate hosts a $|C|=1$ quantized anomalous Hall (QAH) insulator, and a whole series of fractional quantum anomalous Hall (FQAH) states that spontaneously break time-reversal symmetry.

Finding Chern bands that are favorable to realize such fractional states is an ongoing challenge~\cite{PhysRevX.1.021014,Sheng:2011tr,NeupertFQHZeroField,zhao_review,Neupertreview_2015,Liu_review_2023}. In initial realizations of fractional Chern insulators (FCI), magnetic field was necessary to induce~\cite{SpantonFCI}, or improve the Chern band~\cite{FCI_TBG_exp,parker2021field} in order to favor fractional states. A zero-field fractional quantum Hall effect was only recently realized in \ce{MoTe2}~\cite{cai2023signatures, zeng2023thermodynamic, park2023observation, xu2023observation}. For all of the aforementioned systems, the Chern band has been understood at a single-particle level, with interactions providing isospin polarization and thence the fractional states~\cite{bultinck2020mechanism, xie2020nature,zou2018band,zhang2019twisted, wu2019topological,wang2023topological}. In contrast, the origin and the character of the Chern band in rhombohedral pentalayer graphene~\cite{lu2023fractional} has so far been a mystery.

In this work, we shed light on this mystery by numerically solving the interacting many-body problem in $N_L = 4-6$ layer RMG with an hBN potential. Using self-consistent Hartree-Fock (SCHF) calculations, we find that interactions open a gap at filling $\nu=1$, stabilizing an isolated $\n{C}=1$ Chern band. Using exact diagonalization (ED) and density-matrix renormalization group (DMRG) calculations, we confirm that this Chern band hosts an FQAH state.

\begin{figure}[t]
	\includegraphics[width=\linewidth]{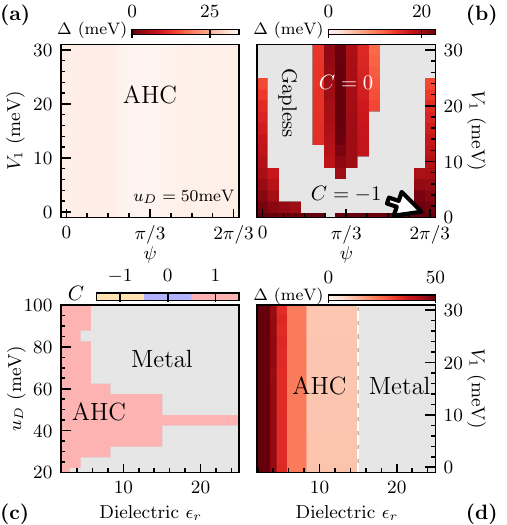}
	\caption{
    (a) The featureless charge gap of rhombohedral pentalayer graphene at $\nu=1$ on the moir\'{e}-distant side ($u_D = \SI{50}{meV}$) as the moir\'e strength $V_1$ and phase $\psi$ vary.
    (b) Same on the moir\'e-proximate side ($u_D = -\SI{50}{meV}$), with a trivial insulator separated from a small Chern insulating phase by a metallic region.
    (c) The Chern number with respect to $u_D$ and dielectric constant $\epsilon_r$ on the moir\'{e}-distant side, showing the AHC survives to relatively weak interaction strengths near $u_D \sim{} \SI{50}{meV}$.
    (d) The charge gap at $u_D = \SI{50}{meV}$ versus moir\'e strength $V_1$ and dielectric $\epsilon_r$. The AHC phase reaches $V_1 = 0$, demonstrating interaction-driven spontaneous translation symmetry breaking.
    All data from SCHF at $\theta=0.77^\circ$.
    }
	\label{fig:robustness}
\end{figure}

The Chern band is remarkably robust to perturbations. Crucially, it arises when the electrons are polarized \textit{away} from the aligned hBN substrate, which we refer to as the ``moir\'e-distant" case (Fig.~\ref{fig:intro}(a)). It follows that the Chern band is insensitive to many details; similar Chern bands arise for different moir\'{e} potentials, different hopping parameters, and for $N_L=4-6$ rhombohedral layers (Fig.~\ref{fig:intro}(c)).
In fact, the Chern band persists even as the moir\'e potential is turned off within our SCHF calculations (Fig.~\ref{fig:intro}(b), \ref{fig:robustness}(a)(d)). This is reminiscent of the ``Hall crystal'' phase proposed in~\cite{hallcrystal1986PRL,hallcrystal1989PRB}, where electrons in a magnetic field crystallize into a state with nonzero Hall conductance. Thus, we propose that the Chern insulator at $\nu=1$ is an anomalous Hall crystal (AHC) state that spontaneously breaks both continuous translation symmetry and time reversal symmetry (with the moir\'e potential providing a pinning field). This establishes an entirely new mechanism for Chern band formation that arises from the interplay of translation breaking and topology.

\begin{figure*}[t]
    \includegraphics[width=\textwidth]{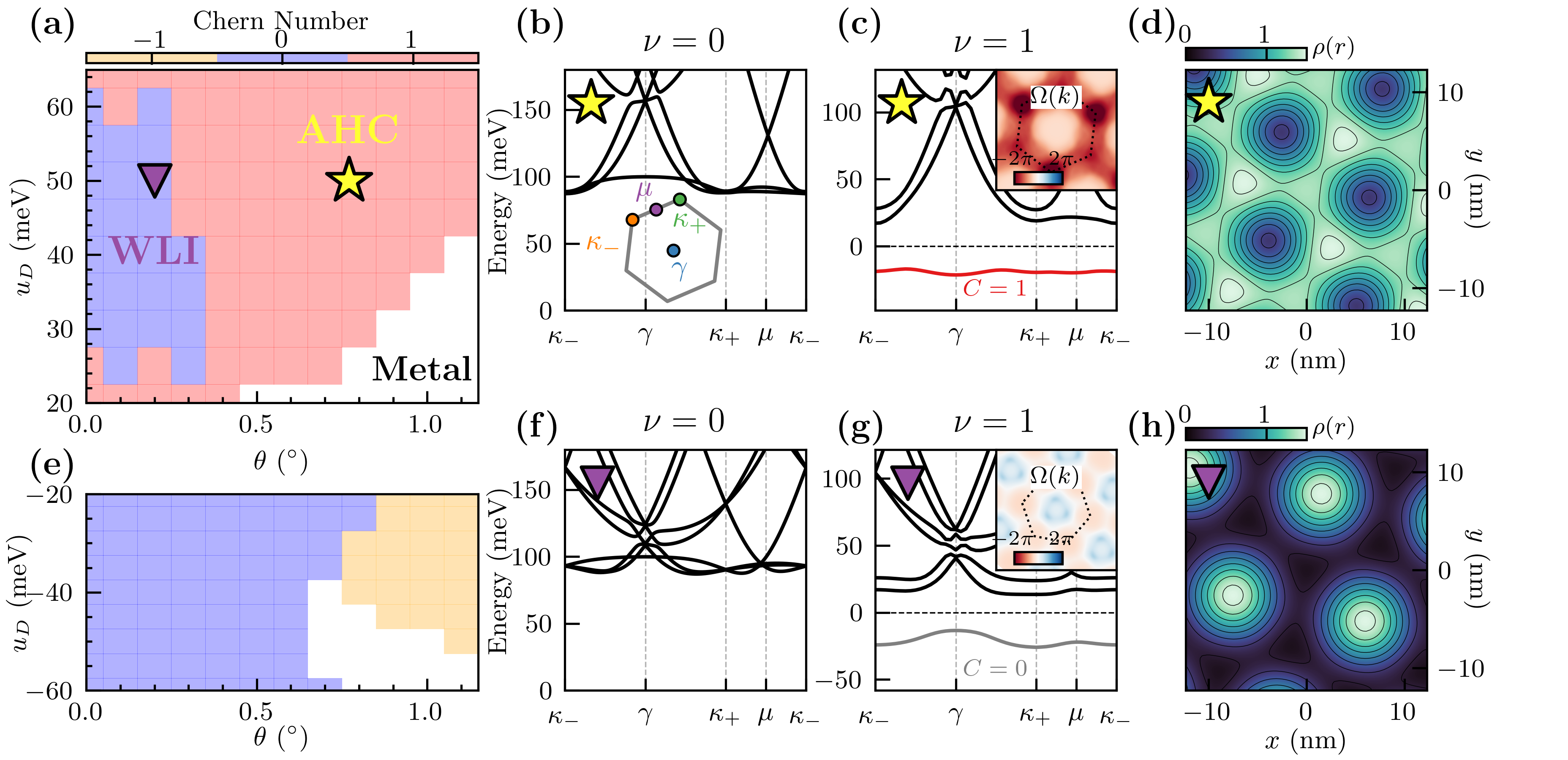}
    \caption{(a,e) The phase diagram of moir\'e rhombohedral pentalayer graphene as a function of the interlayer potential difference $u_D$, and the twist angle $\theta$. The colors label the Chern number of the SCHF ground states, with gapless regions shown in white. On the moir\'e-distant side ($u_D > 0$), the anomalous Hall crystal (AHC) phase has $C=1$, whereas the Wigner-like insulator (WLI) has $C=0$. 
    (b,f) Non-interacting band structures at the yellow star $(u_D, \theta)=(\SI{50}{meV}, 0.77^\circ)$ and purple triangle $(u_D, \theta)=(\SI{50}{meV}, 0.2^\circ)$. Inset in (b): plot of the moir\'e Brillouin zone.
    (c,g) SCHF bandstructure at $\nu=1$, finding an AHC and WLI at the yellow star and purple triangle respectively. Insets: Berry curvature $\Omega(\v k)$ of the occupied band.
    (d,h) Charge densities $\rho(\v r)$ of the AHC (d) and WLI (h). The AHC charge density resembles a honeycomb lattice, whereas the WLI charge density resembles a triangular lattice.
    }
	\label{fig:phase_details}
\end{figure*}


\PRLsec{Microscopic Model} We consider a model~\cite{zhang2010band,jung2013gapped} of RMG with a moir\'e potential~\cite{moon2014electronic} due to an hBN substrate at twist angle $\theta$ (Fig.~\ref{fig:intro}(a)): 
\begin{equation}
    \hat{H} = \hat{h}_{\textrm{kin}}+\sum_{\v{q}}\frac{U_{|\v q|}}{2A} :\hat{\rho}_{\v q}\hat{\rho}_{-\v q}:,
    \quad U_{q} = \frac{2\pi \tanh(qd)}{\epsilon_r\epsilon_0 q},
\end{equation}
where $\hat{\rho}_{\v q}$ is the density operator at momentum $\v{q}$, $A$ is the sample area, normal ordering is relative to the charge neutrality gap of \RNG{} in the presence of displacement field, and the screened Coulomb potential $U_{|\v{q}|}$ has gate distance $d = \SI{250}{\angstrom}$ and dielectric constant $\epsilon_r = 5$.

The kinetic term in the $K$-valley takes the form
\begin{equation}
    h_\textrm{kin} = h_{RG}^{(N_L)} + h_{D} + V_{\mathrm{hBN}},
\end{equation}
where $h_{RG}^{(N_L)}$ is a standard model~\cite{zhou2021half} for $N_L$-layer rhombohedral graphene with in-layer and interlayer hoppings $t_{0-4}$.
The displacement term $[h_D]_{\ell \ell'} = u_D(\ell-1-(N_L-1)/2) \delta_{\ell \ell'}$ creates a potential difference of $u_D$ per layer with the zero point in the middle. Physically, $u_D > 0$ polarizes electrons in the conduction band \textit{away} from the hBN moir\'e potential. The potential acts on the bottom layer as $V_{\mathrm{hBN}}(\v{r}) = V_0 \sigma^0 + V_1 \v{f}(\v{r}, \psi) \cdot \v{\sigma}$, which is a sum of first harmonics with strength $V_1$ and phase $\psi$, plus a layer potential shift $V_0$ ($\sigma^{\mu}$ are sublattice Pauli matrices).
While a variety of models have been proposed for $V_{\mathrm{hBN}}$~\cite{moon2014electronic,JungAbInitio,jung2017moire, krisna2023moire}, our focus is on the moir\'e-distant side $u_D > 0$, which we find is largely $V_{\textrm{hBN}}$-independent. As the displacement field and moir\'e potential break $C_{2x}$ and $M_x$ symmetries of standalone \RNG, the moir\'e Hamiltonian $\hat{H}$ retains only $C_{3z}$ rotation, time-reversal, and moir\'e translation symmetries. See App.~\ref{app:microscopic_model}~\cite{footnoteSM} for complete model details.


\PRLsec{Interaction-Induced Chern Band} To understand the origin of the Chern insulator, we performed self-consistent Hartree-Fock (SCHF) calculations on up to $48\times48$ moir\'e unit cells. We project the $N_L=5$ graphene layers to the lowest $7$ conduction bands, which is well-controlled whenever $\n{u_D}$ is sizable.
The relevant Chern insulator recently observed in rhombohedral pentalayer graphene was at electron filling $\nu = 1$, which suggests a flavor polarization into a single species. Therefore we assume polarization to a single valley and spin flavor.

We first consider $\theta=0.77^\circ$ and $u_D = \SI{50}{\milli\electronvolt}$, estimated to match the displacement field $D/\epsilon_0=\SI{0.81}{\volt/\nano\meter}$ that stabilizes the QAH insulator experimentally~\footnote{We estimate the conversion between $D$ and $u_D$ by comparing the reported HF phase diagram with the experiment in rhombohedral trilayer graphene~\cite{chatterjee2022inter,zhou2021half}.}. The single-particle band structure of moir\'e-distant electrons (Fig.~\ref{fig:phase_details}(b)) \textit{does not} have a direct gap above the first band. Within SCHF at $\nu=1$ (Fig.~\ref{fig:phase_details}(c)), we find interactions open a gap, where the occupied band (red) has strictly positive Berry curvature (inset) and Chern number $\n{C}=1$. 
We refer to this as the AHC phase, for reasons justified below. Its narrow \SI{6}{meV} bandwidth and favorable quantum geometry, discussed below, make it eminently suitable for fractional topological states.


\PRLsec{Phase Diagram at $\nu=1$} Fig.~\ref{fig:phase_details}(a,e) shows the phase diagram of moir\'{e} rhombohedral pentalayer graphene at filling $\nu=1$. On the moir\'e-distant side $u_D > 0$, the AHC phase is stabilized over a broad range of twist angles and displacement fields near the experimentally relevant point (yellow star). Decreasing the displacement field drives a transition into a metallic phase, consistent with experiment~\cite{lu2023fractional}. Reducing the twist angle $\theta$ instead drives a transition to a $C=0$ insulator that we call a ``Wigner-like insulator" (WLI), based on its charge distribution $\rho(\v{r})$. The band structures of a representative WLI state is shown in Fig.~\ref{fig:phase_details}(g). The charge density of the state (Fig.~\ref{fig:phase_details}(h))  shows that $\rho(r)$ varies strongly in the WLI, with peaks on a triangular lattice, reminiscent of a Wigner crystal. Its single-particle band structure in Fig.~\ref{fig:phase_details}(f) reveals the origin of this charge density variation:  at the $\gamma$ point, the backfolded bands decrease in energy as $\theta$ decreases, bringing many within $\SI{30}{\milli\electronvolt}$ of the minimum. These bands may then be strongly mixed by interactions, selecting the configuration that minimizes Hartree energy, i.e. a triangular lattice. Conversely, the AHC has a more uniform charge distribution without distinct peaks (Fig.~\ref{fig:phase_details}(d)). 

We comment briefly on the moir\'e-proximate side $u_D < 0$, where the dominant phase is a $C=0$ insulator. Though consistent with experiment~\cite{lu2023fractional}, we caution that this phase governed by details of $V_{\mathrm{hBN}}$, and Fig. \ref{fig:robustness}(b) shows that varying the moir\'e potential strength and phase can drive a transition to a metal or even a Chern insulator. Models incorporating lattice relaxation~\cite{krisna2023moire}, other \textit{ab initio} modelling~\cite{jung2017moire,park2023topological}, and comparison with experiment~\cite{lee2016ballistic} may be required here.~\footnote{Near $u_D \approx 0$ valence bands, which we ignore, also become relevant; our model is most reliable for the moir\'e-distant regime $u_D > 0$.}

\PRLsec{Robustness of the Anomalous Hall Crystal} The AHC state is robust to a wide range of perturbations. In Fig.~\ref{fig:robustness}(c), we show the Chern number as a function of $u_D$ and dielectric constant $\epsilon_r$. The AHC state survives up to $\epsilon_r \sim 15$ for a range of $u_D$, showing that relatively weak interactions are sufficient to stabilize it. Fig.~\ref{fig:robustness}(a, d) shows the moir\'e-distant charge gap as a function of $\epsilon_r$ and moir\'{e} potential strength $V_1$ and phase $\psi$. The charge gap increases with interaction strength, but is essentially independent of the moir\'e potential. The charge gap is peaked near $u_D \approx \SI{55}{meV}$ at $\theta\approx 0.77^\circ$, despite the effective moir\'e potential decreasing monotonically with $u_D$ (see App.~\ref{app:HartreeFock}~\cite{footnoteSM}). 
Conversely, tuning $V_1$ and $\psi$ changes the ground state on the moir\'e-proximate side (Fig.~\ref{fig:robustness}(b)).

Remarkably, the AHC phase does not require a particular number of rhombohedral layers. Fig.~\ref{fig:intro}(c) shows SCHF ground state at $\nu=1$ for $N_L=4-6$ layers, with an $\n{C}=1$ AHC phase appearing in all cases at large $u_D$. Moreover, the critical $u_D^*$ decreases significantly as the number of layers increases, easing the experimental requirement of a large displacement field $D \propto u_D$. Finally, the phase diagram is robust to modelling differences in the hopping parameters of rhombohedral graphene. For instance, we show in the Supplemental Material that the AHC remains when all further neighbor hoppings $t_{i\geq 2}$ are set to zero.

\PRLsec{Spontaneous translation symmetry breaking} The resilience of the $|C|=1$ phase to the change in moir\'{e} potential raises a natural question: Is the moir\'{e} potential necessary at all for stabilizing the state? We now argue that within SCHF the ground state is an anomalous Hall crystal~\cite{hallcrystal1986PRL, hallcrystal1989PRB} that spontaneously breaks translation symmetry.

We consider the Hamiltonian with \textit{zero} moir\'{e} potential. Under this condition, SCHF calculations at $n\approx 0.93\times 10^{12}\si{cm^{-2}}$ find a ground state with \textit{spontaneously} broken graphene translation symmetry, which has much lower energy than (metallic) graphene-translation symmetric states when $u_D>\SI{25}{\milli\electronvolt}$ (Fig.~\ref{fig:intro}(b)). We note that the AHC state is most stable compared to the metallic phase around $u_D=\SI{55}{\milli\electronvolt}$, which is consistent with our estimate for the experimental displacement field and where the charge gap reaches maximum.

The AHC state is a compressible state in the clean limit, since the lattice constant can change continuously. However, a small spatially-varying potential or disorder can pin its periodicity~\cite{hallcrystal1989PRB}, whereupon its nontrivial Chern number can be revealed through the Str{e}da formula and Hall conductance. This is likely the role played by moir\'{e} potential in experiment~\cite{lu2023fractional}: both the topology and the charge gap of the ground state are determined by the interaction effects that are already present in the moir\'e-less limit, but the moir\'{e} pinning field stabilizes the crystal at $\nu=1$.


\begin{figure}[t]
	\includegraphics[width=\linewidth]{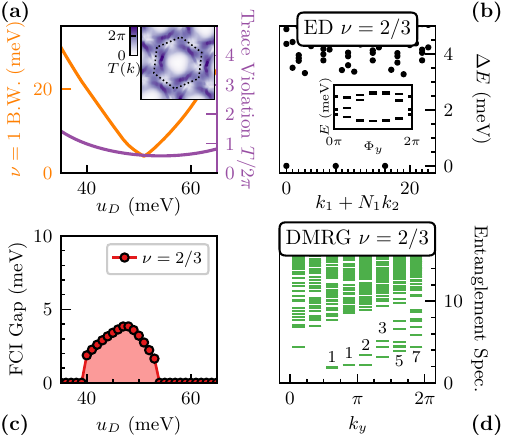}
	\caption{(a) Band conditions of the AHC phase as a function of $u_D$. Both the bandwidth and the violation of trace condition are minimized near $u_D=\SI{50}{\milli\electronvolt}$.
    (b) Many-body spectrum from ED on a $4\times 6$ system at $\nu=2/3$. Inset: spectral flow of the three degenerate ground states, a signature of the FCI ground state~\cite{PhysRevX.1.021014,Sheng:2011tr,NeupertFQHZeroField}.
    (c) The FCI gap at $\nu=2/3$ as a function of $u_D$ obtained in ED. The FCI gap is nonzero around $40$ to $55$ \si{\milli\electronvolt}, and gives way to metals beyond this regime due to large bandwidths.
    (d) Entanglement spectrum obtained in DMRG at $\nu=2/3$ with $L_y=8$. The partition number counting of states below the entanglement gap is consistent with the chiral edge modes associated with the Laughlin states, confirming the state is an FCI~\cite{li2004entanglement}. Parameters: $(u_D, \theta)=(\SI{50}{meV}, 0.77^\circ)$ unless specified.
    }
	\label{fig:FCI}
\end{figure}

    \PRLsec{Fractional quantum anomalous Hall (FQAH) effect} One of the main features of the experimental discovery~\cite{lu2023fractional} was a full series of FQAH states found at fractional fillings in the absence of magnetic field. Employing exact diagonalization (ED) and density matrix renormalization group (DMRG) calculations, here we numerically confirm that partially filling the Chern band generated by the AHC gives us FQAH states.

Recent progress on the study of FCI phases has emphasized the role of quantum geometry~\cite{PhysRevB.90.165139,Parameswaran_2013,Jackson:2015aa,Martin_PositionMomentumDuality,Grisha_TBG,Grisha_TBG2,kahlerband1,kahlerband2,kahlerband3,JieWang_exactlldescription,Jie_hierarchyidealband,LedwithVishwanathParker22,JW_origin_22,junkaidonghighC22,crepel_chiral_2023,valentin23ideal,andrews2023stability}, in addition to flat dispersion~\cite{KaiSunNearlyFlat,NeupertFQHZeroField,WangFQHBoson,TangHighTemperature,Sheng:2011tr,PhysRevX.1.021014}. Small trace condition violation $T = \int d^2\v{k} (\Tr(g_\mathrm{FS}(\v{k})) - \Omega(\v{k}))$ gives a band suitable for finding FCIs. Here, $g_\mathrm{FS}(\v{k})$ is the Fubini-Study metric, and $\Omega(\v{k})$ is the Berry curvature.
The AHC band has both flat bandwidth and relatively small trace condition violation (Fig.~\ref{fig:FCI}(a)), making it likely to host FQAH phases.  Indeed we find that a Laughlin state appears at $\nu=2/3$ of the AHC band, with clear signatures from both ED (Fig.~\ref{fig:FCI}(b)) and DMRG calculations (Fig.~\ref{fig:FCI}(d)). These calculations~\cite{footnoteSM} are done in the many-body Hilbert space of the lowest SCHF band at $\nu=1$, assuming spin and valley polarization, which numerically biases the formation of fractional topological states. The FQAH phase yields to a competing metallic phase at small and large $u_D$ (Fig.~\ref{fig:FCI}(c)). The FQAH state is likely a ground state in the spin- and valley-polarized sector. Similar to the case in the lowest Landau level~\cite{PhysRevB.41.7910,PhysRevLett.72.3405,PhysRevB.53.15845}, this state is likely in competition with a spin-unpolarized state, which merits future investigation.

Examining the stability of the FQAH phases down to the zero moir\'e potential limit is an excellent topic for future studies, where we expect competition between an FQAH phase with the same unit cell size, versus Wigner or anomalous Hall crystals with larger unit cells.
Experimentally, the existence of the $\nu=2/3$ plateau points to the realization of the former scenario, at least in the weak moir\'{e} potential limit, while a profusion of insulating states at smaller densities~\cite{lu2023fractional} might be due to the latter scenario.

\PRLsec{Discussion} We have demonstrated the existence of the AHC phase within SCHF calculations. While such a phase can be stabilized \textit{in principle}, SCHF is known to overestimate the tendency to crystallize, for example in Wigner crystals~\cite{Trail_2003}. Pinpointing the location of the phase boundary between the AHC and the metal with more sophisticated techniques (just as variational Monte Carlo is used in the jellium model~\cite{tanatar1989, attaccalite2002, drummond2009}) is left to future work. We note that the AHC phase at the optimal $u_D$ survives even after decreasing the interaction strength by a factor of three (Fig.~\ref{fig:robustness}(c,d)), hinting at the robustness of the phase beyond mean-field theory.

The interpretation of the Chern state as an AHC clarifies an
experimental mystery. Experimental phase diagrams in~\cite{lu2023fractional} are strongly asymmetric between positive and negative displacement fields $D$, suggesting that the hBN is only aligned on one side (Fig.~\ref{fig:intro}(a)). Comparing experimental data at $\nu=4$~\cite{lu2023fractional} to SCHF calculations of the charge gap~\cite{footnoteSM}, we independently infer that Chern insulator appears when the electrons are polarized \text{away} from the hBN potential (moir\'{e}-distant) instead of towards the hBN potential (moir\'{e}-proximate). We also predict that AHC in samples without moir\'e would have a characteristic density pattern and $\sigma^{xy} = e^2/h$ over extended density ranges.

The ubiquity of the AHC phase in our phase diagram hints at a universal physical mechanism stabilizing the formation of a Chern band, while the relatively uniform charge density of the AHC state points to a translation-breaking mechanism different from that of the Wigner crystal. 
A deeper analytic understanding of the origin of the AHC phase is clearly needed, which will be addressed in a forthcoming work~\cite{soejima2024anomaloushallcrystalsrhombohedral}. 

\PRLsec{Note added} Towards the completion of the project, four interesting related papers~\cite{dong2023theory, zhou2023fractional,guo_theory_2023,kwan_moire_2023} were posted, which agree with our finding in the areas of overlap.

\begin{acknowledgements}
We acknowledge Long Ju, Tonghang Han, Jixiang Yang, Yves Kwan, Patrick J. Ledwith, Eslam Khalaf, Trithep Devakul, and Bertrand I. Halperin for fruitful and insightful discussions. T.W., T.W., and M.Z. are supported by the U.S. Department of Energy, Office of Science, Office of Basic Energy Sciences, Materials Sciences and Engineering Division under Contract No. DE-AC02-05-CH11231 (Theory of Materials program KC2301). A.V. is supported by the Simons Collaboration on UltraQuantum Matter, which is a grant from the Simons
Foundation (651440, A.V.) and by the Center for Advancement of Topological Semimetals, an Energy Frontier Research Center funded by the US Department of
Energy Office of Science, Office of Basic Energy Sciences,
through the Ames Laboratory under contract No. DEAC02-07CH11358. This research used the Lawrencium computational cluster provided by the Lawrence Berkeley National Laboratory (Supported by the U.S. Department of Energy, Office of Basic Energy Sciences under Contract No. DE-AC02-05-CH11231). This research is funded in part by the
Gordon and Betty Moore Foundation’s EPiQS Initiative,
Grant GBMF8683 to T.S. D.E.P. is supported by the Simons Collaboration on Ultra-Quantum Matter, which is a grant from the Simons Foundation (1151944, MPZ).
\end{acknowledgements}

\bibliographystyle{unsrt}
\bibliography{bibliography, bibliography_TBG,CFL_bib}

\begin{thebibliography}{100}

\bibitem{lu2023fractional}
Zhengguang Lu, Tonghang Han, Yuxuan Yao, Aidan~P Reddy, Jixiang Yang, Junseok Seo, Kenji Watanabe, Takashi Taniguchi, Liang Fu, and Long Ju.
\newblock Fractional quantum anomalous hall effect in a graphene moire superlattice.
\newblock {\em arXiv preprint arXiv:2309.17436}, 2023.

\bibitem{xie2021fractional}
Yonglong Xie, Andrew~T. Pierce, Jeong~Min Park, Daniel~E. Parker, Eslam Khalaf, Patrick Ledwith, Yuan Cao, Seung~Hwan Lee, Shaowen Chen, Patrick~R. Forrester, Kenji Watanabe, Takashi Taniguchi, Ashvin Vishwanath, Pablo Jarillo-Herrero, and Amir Yacoby.
\newblock Fractional chern insulators in magic-angle twisted bilayer graphene.
\newblock {\em Nature}, 600(7889):439--443, dec 2021.

\bibitem{cao2018correlated}
Yuan Cao, Valla Fatemi, Ahmet Demir, Shiang Fang, Spencer~L. Tomarken, Jason~Y. Luo, Javier~D. Sanchez-Yamagishi, Kenji Watanabe, Takashi Taniguchi, Efthimios Kaxiras, Ray~C. Ashoori, and Pablo Jarillo-Herrero.
\newblock Correlated insulator behaviour at half-filling in magic-angle graphene superlattices.
\newblock {\em Nature}, 556(7699):80--84, mar 2018.

\bibitem{Cory_Science19}
Matthew Yankowitz, Shaowen Chen, Hryhoriy Polshyn, Yuxuan Zhang, K.~Watanabe, T.~Taniguchi, David Graf, Andrea~F. Young, and Cory~R. Dean.
\newblock Tuning superconductivity in twisted bilayer graphene.
\newblock {\em Science}, 363(6431):1059--1064, 2019.

\bibitem{Efetov2020Screening}
Petr Stepanov, Ipsita Das, Xiaobo Lu, Ali Fahimniya, Kenji Watanabe, Takashi Taniguchi, Frank H.~L. Koppens, Johannes Lischner, Leonid Levitov, and Dmitri~K. Efetov.
\newblock {Untying the insulating and superconducting orders in magic-angle graphene}.
\newblock {\em Nature}, 583(7816):375--378, July 2020.

\bibitem{jaoui_quantum_2022}
Alexandre Jaoui, Ipsita Das, Giorgio Di~Battista, Jaime Díez-Mérida, Xiaobo Lu, Kenji Watanabe, Takashi Taniguchi, Hiroaki Ishizuka, Leonid Levitov, and Dmitri~K. Efetov.
\newblock Quantum critical behaviour in magic-angle twisted bilayer graphene.
\newblock {\em Nature Physics}, April 2022.

\bibitem{stepanov_competing_2021}
Petr Stepanov, Ming Xie, Takashi Taniguchi, Kenji Watanabe, Xiaobo Lu, Allan~H. MacDonald, B.~Andrei Bernevig, and Dmitri~K. Efetov.
\newblock Competing zero-field chern insulators in superconducting twisted bilayer graphene.
\newblock {\em Physical Review Letters}, 127(19):197701, November 2021.

\bibitem{cao2018unconventional}
Yuan Cao, Valla Fatemi, Shiang Fang, Kenji Watanabe, Takashi Taniguchi, Efthimios Kaxiras, and Pablo Jarillo-Herrero.
\newblock Unconventional superconductivity in magic-angle graphene superlattices.
\newblock {\em Nature}, 556(7699):43--50, mar 2018.

\bibitem{zondiner_cascade_2020}
U.~Zondiner, A.~Rozen, D.~Rodan-Legrain, Y.~Cao, R.~Queiroz, T.~Taniguchi, K.~Watanabe, Y.~Oreg, F.~von Oppen, Ady Stern, E.~Berg, P.~Jarillo-Herrero, and S.~Ilani.
\newblock Cascade of phase transitions and {Dirac} revivals in magic-angle graphene.
\newblock {\em Nature}, 582(7811):203--208, June 2020.

\bibitem{park_flavour_2021}
Jeong~Min Park, Yuan Cao, Kenji Watanabe, Takashi Taniguchi, and Pablo Jarillo-Herrero.
\newblock Flavour {Hund}’s coupling, {Chern} gaps and charge diffusivity in moiré graphene.
\newblock {\em Nature}, 592(7852):43--48, April 2021.

\bibitem{cao2020_nematicity}
Yuan Cao, Daniel Rodan-Legrain, Jeong~Min Park, Noah F.~Q. Yuan, Kenji Watanabe, Takashi Taniguchi, Rafael~M. Fernandes, Liang Fu, and Pablo Jarillo-Herrero.
\newblock {Nematicity and competing orders in superconducting magic-angle graphene}.
\newblock {\em Science}, 372(6539):264--271, 2021.

\bibitem{Yankowitz2019}
Matthew Yankowitz, Shaowen Chen, Hryhoriy Polshyn, Yuxuan Zhang, K.~Watanabe, T.~Taniguchi, David Graf, Andrea~F. Young, and Cory~R. Dean.
\newblock {Tuning superconductivity in twisted bilayer graphene}.
\newblock {\em Science}, 363(6431):1059--1064, 2019.

\bibitem{saito_independent_2020}
Yu~Saito, Jingyuan Ge, Kenji Watanabe, Takashi Taniguchi, and Andrea~F. Young.
\newblock Independent superconductors and correlated insulators in twisted bilayer graphene.
\newblock {\em Nature Physics}, 16(9):926--930, September 2020.

\bibitem{wong_cascade_2020}
Dillon Wong, Kevin~P. Nuckolls, Myungchul Oh, Biao Lian, Yonglong Xie, Sangjun Jeon, Kenji Watanabe, Takashi Taniguchi, B.~Andrei Bernevig, and Ali Yazdani.
\newblock Cascade of electronic transitions in magic-angle twisted bilayer graphene.
\newblock {\em Nature}, 582(7811):198--202, June 2020.

\bibitem{oh_evidence_2021}
Myungchul Oh, Kevin~P. Nuckolls, Dillon Wong, Ryan~L. Lee, Xiaomeng Liu, Kenji Watanabe, Takashi Taniguchi, and Ali Yazdani.
\newblock Evidence for unconventional superconductivity in twisted bilayer graphene.
\newblock {\em Nature}, 600(7888):240--245, December 2021.

\bibitem{choi2021_STM}
Youngjoon Choi, Hyunjin Kim, Cyprian Lewandowski, Yang Peng, Alex Thomson, Robert Polski, Yiran Zhang, Kenji Watanabe, Takashi Taniguchi, Jason Alicea, and Stevan Nadj-Perge.
\newblock {Interaction-driven band flattening and correlated phases in twisted bilayer graphene}.
\newblock {\em Nat. Phys.}, 17:1375--1381, 2021.

\bibitem{yu_correlated_2022}
Jiachen Yu, Benjamin~A. Foutty, Zhaoyu Han, Mark~E. Barber, Yoni Schattner, Kenji Watanabe, Takashi Taniguchi, Philip Phillips, Zhi-Xun Shen, Steven~A. Kivelson, and Benjamin~E. Feldman.
\newblock Correlated {Hofstadter} spectrum and flavour phase diagram in magic-angle twisted bilayer graphene.
\newblock {\em Nature Physics}, April 2022.

\bibitem{chen2020monobi}
Shaowen Chen, Minhao He, Ya-Hui Zhang, Valerie Hsieh, Zaiyao Fei, K.~Watanabe, T.~Taniguchi, David~H. Cobden, Xiaodong Xu, Cory~R. Dean, and Matthew Yankowitz.
\newblock Electrically tunable correlated and topological states in twisted monolayer{\textendash}bilayer graphene.
\newblock {\em Nature Physics}, 17(3):374--380, oct 2020.

\bibitem{polshyn2020electrical}
H.~Polshyn, J.~Zhu, M.~A. Kumar, Y.~Zhang, F.~Yang, C.~L. Tschirhart, M.~Serlin, K.~Watanabe, T.~Taniguchi, A.~H. MacDonald, and A.~F. Young.
\newblock Electrical switching of magnetic order in an orbital chern insulator.
\newblock {\em Nature}, 588(7836):66--70, nov 2020.

\bibitem{polshyn2021topological}
H.~Polshyn, Y.~Zhang, M.~A. Kumar, T.~Soejima, P.~Ledwith, K.~Watanabe, T.~Taniguchi, A.~Vishwanath, M.~P. Zaletel, and A.~F. Young.
\newblock Topological charge density waves at half-integer filling of a moir{\'{e}} superlattice.
\newblock {\em Nature Physics}, 18(1):42--47, dec 2021.

\bibitem{zhang2023local}
Canxun Zhang, Tiancong Zhu, Tomohiro Soejima, Salman Kahn, Kenji Watanabe, Takashi Taniguchi, Alex Zettl, Feng Wang, Michael~P. Zaletel, and Michael~F. Crommie.
\newblock Local spectroscopy of a gate-switchable moir{\'{e}} quantum anomalous hall insulator.
\newblock {\em Nature Communications}, 14(1), jun 2023.

\bibitem{cao2020doublebilayer}
Yuan Cao, Daniel Rodan-Legrain, Oriol Rubies-Bigorda, Jeong~Min Park, Kenji Watanabe, Takashi Taniguchi, and Pablo Jarillo-Herrero.
\newblock Tunable correlated states and spin-polarized phases in twisted bilayer{\textendash}bilayer graphene.
\newblock {\em Nature}, 583(7815):215--220, may 2020.

\bibitem{burg2019correlated}
G.~William Burg, Jihang Zhu, Takashi Taniguchi, Kenji Watanabe, Allan~H. MacDonald, and Emanuel Tutuc.
\newblock Correlated insulating states in twisted double bilayer graphene.
\newblock {\em Physical Review Letters}, 123(19), nov 2019.

\bibitem{liu2020tunable}
Xiaomeng Liu, Zeyu Hao, Eslam Khalaf, Jong~Yeon Lee, Yuval Ronen, Hyobin Yoo, Danial~Haei Najafabadi, Kenji Watanabe, Takashi Taniguchi, Ashvin Vishwanath, and Philip Kim.
\newblock Tunable spin-polarized correlated states in twisted double bilayer graphene.
\newblock {\em Nature}, 583(7815):221--225, jul 2020.

\bibitem{he2021symmetry}
Minhao He, Yuhao Li, Jiaqi Cai, Yang Liu, K~Watanabe, Takashi Taniguchi, Xiaodong Xu, and Matthew Yankowitz.
\newblock Symmetry breaking in twisted double bilayer graphene.
\newblock {\em Nature Physics}, 17(1):26--30, 2021.

\bibitem{shen2020correlated}
Cheng Shen, Yanbang Chu, QuanSheng Wu, Na~Li, Shuopei Wang, Yanchong Zhao, Jian Tang, Jieying Liu, Jinpeng Tian, Kenji Watanabe, et~al.
\newblock Correlated states in twisted double bilayer graphene.
\newblock {\em Nature Physics}, 16(5):520--525, 2020.

\bibitem{hao2021electric}
Zeyu Hao, A.~M. Zimmerman, Patrick Ledwith, Eslam Khalaf, Danial~Haie Najafabadi, Kenji Watanabe, Takashi Taniguchi, Ashvin Vishwanath, and Philip Kim.
\newblock Electric field{\textendash}tunable superconductivity in alternating-twist magic-angle trilayer graphene.
\newblock {\em Science}, 371(6534):1133--1138, mar 2021.

\bibitem{park2021tunable}
Jeong~Min Park, Yuan Cao, Kenji Watanabe, Takashi Taniguchi, and Pablo Jarillo-Herrero.
\newblock Tunable strongly coupled superconductivity in magic-angle twisted trilayer graphene.
\newblock {\em Nature}, 590(7845):249--255, feb 2021.

\bibitem{park2022robust}
Jeong~Min Park, Yuan Cao, Li-Qiao Xia, Shuwen Sun, Kenji Watanabe, Takashi Taniguchi, and Pablo Jarillo-Herrero.
\newblock Robust superconductivity in magic-angle multilayer graphene family.
\newblock {\em Nature Materials}, 21(8):877--883, jul 2022.

\bibitem{xia2023helical}
Li-Qiao Xia, Sergio~C. de~la Barrera, Aviram Uri, Aaron Sharpe, Yves~H. Kwan, Ziyan Zhu, Kenji Watanabe, Takashi Taniguchi, David Goldhaber-Gordon, Liang Fu, Trithep Devakul, and Pablo Jarillo-Herrero.
\newblock Helical trilayer graphene: a moir\'e platform for strongly-interacting topological bands, 2023.

\bibitem{uri2023quasicrystal}
Aviram Uri, Sergio~C. de~la Barrera, Mallika~T. Randeria, Daniel Rodan-Legrain, Trithep Devakul, Philip J.~D. Crowley, Nisarga Paul, Kenji Watanabe, Takashi Taniguchi, Ron Lifshitz, Liang Fu, Raymond~C. Ashoori, and Pablo Jarillo-Herrero.
\newblock Superconductivity and strong interactions in a tunable moir{\'{e}} quasicrystal.
\newblock {\em Nature}, 620(7975):762--767, jul 2023.

\bibitem{zhou2021superconductivity}
Haoxin Zhou, Tian Xie, Takashi Taniguchi, Kenji Watanabe, and Andrea~F Young.
\newblock Superconductivity in rhombohedral trilayer graphene.
\newblock {\em Nature}, 598(7881):434--438, 2021.

\bibitem{Li_2021_2}
Tingxin Li, Shengwei Jiang, Bowen Shen, Yang Zhang, Lizhong Li, Zui Tao, Trithep Devakul, Kenji Watanabe, Takashi Taniguchi, Liang Fu, Jie Shan, and Kin~Fai Mak.
\newblock Quantum anomalous hall effect from intertwined moir{\'{e}} bands.
\newblock {\em Nature}, 600(7890):641--646, dec 2021.

\bibitem{zhao2022realization}
Wenjin Zhao, Kaifei Kang, Lizhong Li, Charles Tschirhart, Evgeny Redekop, Kenji Watanabe, Takashi Taniguchi, Andrea Young, Jie Shan, and Kin~Fai Mak.
\newblock Realization of the haldane chern insulator in a moir\'e lattice, 2022.

\bibitem{tao_valley-coherent_2022}
Zui Tao, Bowen Shen, Shengwei Jiang, Tingxin Li, Lizhong Li, Liguo Ma, Wenjin Zhao, Jenny Hu, Kateryna Pistunova, Kenji Watanabe, Takashi Taniguchi, Tony~F. Heinz, Kin~Fai Mak, and Jie Shan.
\newblock Valley-coherent quantum anomalous {Hall} state in {AB}-stacked {MoTe2}/{WSe2} bilayers, August 2022.
\newblock arXiv:2208.07452 [cond-mat].

\bibitem{foutty2023mapping}
Benjamin~A. Foutty, Carlos~R. Kometter, Trithep Devakul, Aidan~P. Reddy, Kenji Watanabe, Takashi Taniguchi, Liang Fu, and Benjamin~E. Feldman.
\newblock Mapping twist-tuned multi-band topology in bilayer wse$_2$, 2023.

\bibitem{cai2023signatures}
Jiaqi Cai, Eric Anderson, Chong Wang, Xiaowei Zhang, Xiaoyu Liu, William Holtzmann, Yinong Zhang, Fengren Fan, Takashi Taniguchi, Kenji Watanabe, et~al.
\newblock Signatures of fractional quantum anomalous hall states in twisted mote2.
\newblock {\em Nature}, pages 1--3, 2023.

\bibitem{zeng2023thermodynamic}
Yihang Zeng, Zhengchao Xia, Kaifei Kang, Jiacheng Zhu, Patrick Kn{\"u}ppel, Chirag Vaswani, Kenji Watanabe, Takashi Taniguchi, Kin~Fai Mak, and Jie Shan.
\newblock Thermodynamic evidence of fractional chern insulator in moir{\'e} mote2.
\newblock {\em Nature}, pages 1--2, 2023.

\bibitem{park2023observation}
Heonjoon Park, Jiaqi Cai, Eric Anderson, Yinong Zhang, Jiayi Zhu, Xiaoyu Liu, Chong Wang, William Holtzmann, Chaowei Hu, Zhaoyu Liu, et~al.
\newblock Observation of fractionally quantized anomalous hall effect.
\newblock {\em Nature}, pages 1--3, 2023.

\bibitem{xu2023observation}
Fan Xu, Zheng Sun, Tongtong Jia, Chang Liu, Cheng Xu, Chushan Li, Yu~Gu, Kenji Watanabe, Takashi Taniguchi, Bingbing Tong, et~al.
\newblock Observation of integer and fractional quantum anomalous hall effects in twisted bilayer mote 2.
\newblock {\em Physical Review X}, 13(3):031037, 2023.

\bibitem{anderson2023programming}
Eric Anderson, Feng-Ren Fan, Jiaqi Cai, William Holtzmann, Takashi Taniguchi, Kenji Watanabe, Di~Xiao, Wang Yao, and Xiaodong Xu.
\newblock Programming correlated magnetic states with gate-controlled moir{\'e} geometry.
\newblock {\em Science}, page eadg4268, 2023.

\bibitem{wu_topological_2019}
Fengcheng Wu, Timothy Lovorn, Emanuel Tutuc, Ivar Martin, and A.~H. MacDonald.
\newblock Topological {Insulators} in {Twisted} {Transition} {Metal} {Dichalcogenide} {Homobilayers}.
\newblock {\em Phys. Rev. Lett.}, 122(8):086402, February 2019.
\newblock Publisher: American Physical Society.

\bibitem{yu_giant_2020}
Hongyi Yu, Mingxing Chen, and Wang Yao.
\newblock Giant magnetic field from moir{\textbackslash}'e induced {Berry} phase in homobilayer semiconductors.
\newblock {\em National Science Review}, 7(1):12--20, January 2020.
\newblock arXiv:1906.05499 [cond-mat].

\bibitem{zhai_theory_2020}
Dawei Zhai and Wang Yao.
\newblock Theory of tunable flux lattices in the homobilayer moir{\textbackslash}'e of twisted and uniformly strained transition metal dichalcogenides.
\newblock {\em Phys. Rev. Mater.}, 4(9):094002, September 2020.
\newblock Publisher: American Physical Society.

\bibitem{tang_geometric_2021}
Hao Tang, Stephen Carr, and Efthimios Kaxiras.
\newblock Geometric origins of topological insulation in twisted layered semiconductors.
\newblock {\em Phys. Rev. B}, 104(15):155415, October 2021.
\newblock Publisher: American Physical Society.

\bibitem{devakul_magic_2021}
Trithep Devakul, Valentin Cr{\'e}pel, Yang Zhang, and Liang Fu.
\newblock Magic in twisted transition metal dichalcogenide bilayers.
\newblock {\em Nat Commun}, 12(1):6730, November 2021.
\newblock Number: 1 Publisher: Nature Publishing Group.

\bibitem{zhang_spin-textured_2021}
Yang Zhang, Trithep Devakul, and Liang Fu.
\newblock Spin-textured {Chern} bands in {AB}-stacked transition metal dichalcogenide bilayers.
\newblock {\em Proc. Natl. Acad. Sci. U.S.A.}, 118(36):e2112673118, September 2021.

\bibitem{wang_staggered_2023}
Jie Wang, Jiawei Zang, Jennifer Cano, and Andrew~J. Millis.
\newblock Staggered pseudo magnetic field in twisted transition metal dichalcogenides: {Physical} origin and experimental consequences.
\newblock {\em Phys. Rev. Res.}, 5(1):L012005, January 2023.
\newblock Publisher: American Physical Society.

\bibitem{pan_band_2020}
Haining Pan, Fengcheng Wu, and Sankar Das~Sarma.
\newblock Band topology, {Hubbard} model, {Heisenberg} model, and {Dzyaloshinskii}-{Moriya} interaction in twisted bilayer \$\{{\textbackslash}mathrm\{{WSe}\}\}\_\{2\}\$.
\newblock {\em Phys. Rev. Res.}, 2(3):033087, July 2020.
\newblock Publisher: American Physical Society.

\bibitem{abouelkomsan_multiferroicity_2022}
Ahmed Abouelkomsan, Emil~J. Bergholtz, and Shubhayu Chatterjee.
\newblock Multiferroicity and {Topology} in {Twisted} {Transition} {Metal} {Dichalcogenides}, November 2022.
\newblock arXiv:2210.14918 [cond-mat].

\bibitem{crepel_chiral_2023}
Valentin Cr{\'e}pel, Nicolas Regnault, and Raquel Queiroz.
\newblock The chiral limits of moir{\textbackslash}'e semiconductors: origin of flat bands and topology in twisted transition metal dichalcogenides homobilayers, July 2023.
\newblock arXiv:2305.10477 [cond-mat].

\bibitem{li_electrically_2023}
Bohao Li, Wen-Xuan Qiu, and Fengcheng Wu.
\newblock Electrically tuned topology and magnetism in twisted bilayer {MoTe}\$\_2\$ at \${\textbackslash}nu\_h=1\$, October 2023.
\newblock arXiv:2310.02217 [cond-mat].

\bibitem{qiu_interaction-driven_2023}
Wen-Xuan Qiu, Bohao Li, Xun-Jiang Luo, and Fengcheng Wu.
\newblock Interaction-driven topological phase diagram of twisted bilayer {MoTe}\$\_2\$, October 2023.
\newblock arXiv:2305.01006 [cond-mat].

\bibitem{li_spontaneous_2021}
Heqiu Li, Umesh Kumar, Kai Sun, and Shi-Zeng Lin.
\newblock Spontaneous fractional {Chern} insulators in transition metal dichalcogenide moir{\textbackslash}'e superlattices.
\newblock {\em Phys. Rev. Res.}, 3(3):L032070, September 2021.
\newblock Publisher: American Physical Society.

\bibitem{crepel_anomalous_2023}
Valentin Cr{\'e}pel and Liang Fu.
\newblock Anomalous {Hall} metal and fractional {Chern} insulator in twisted transition metal dichalcogenides.
\newblock {\em Phys. Rev. B}, 107(20):L201109, May 2023.
\newblock Publisher: American Physical Society.

\bibitem{morales-duran_pressure-enhanced_2023}
Nicol{\'a}s Morales-Dur{\'a}n, Jie Wang, Gabriel~R. Schleder, Mattia Angeli, Ziyan Zhu, Efthimios Kaxiras, C{\'e}cile Repellin, and Jennifer Cano.
\newblock Pressure-enhanced fractional {Chern} insulators along a magic line in moir{\textbackslash}'e transition metal dichalcogenides.
\newblock {\em Phys. Rev. Res.}, 5(3):L032022, August 2023.
\newblock Publisher: American Physical Society.

\bibitem{wang_fractional_2023}
Chong Wang, Xiao-Wei Zhang, Xiaoyu Liu, Yuchi He, Xiaodong Xu, Ying Ran, Ting Cao, and Di~Xiao.
\newblock Fractional {Chern} {Insulator} in {Twisted} {Bilayer} {MoTe}\$\_2\$, April 2023.
\newblock arXiv:2304.11864 [cond-mat].

\bibitem{reddy_fractional_2023}
Aidan~P. Reddy, Faisal Alsallom, Yang Zhang, Trithep Devakul, and Liang Fu.
\newblock Fractional quantum anomalous {Hall} states in twisted bilayer \$\{{\textbackslash}mathrm\{{MoTe}\}\}\_\{2\}\$ and \$\{{\textbackslash}mathrm\{{WSe}\}\}\_\{2\}\$.
\newblock {\em Phys. Rev. B}, 108(8):085117, August 2023.
\newblock Publisher: American Physical Society.

\bibitem{jia_moire_2023}
Yujin Jia, Jiabin Yu, Jiaxuan Liu, Jonah Herzog-Arbeitman, Ziyue Qi, Nicolas Regnault, Hongming Weng, B.~Andrei Bernevig, and Quansheng Wu.
\newblock Moir{\textbackslash}'e {Fractional} {Chern} {Insulators} {I}: {First}-principles calculations and {Continuum} {Models} of {Twisted} {Bilayer} {MoTe}\$\_2\$, November 2023.
\newblock arXiv:2311.04958 [cond-mat].

\bibitem{yu_fractional_2023}
Jiabin Yu, Jonah Herzog-Arbeitman, Minxuan Wang, Oskar Vafek, B.~Andrei Bernevig, and Nicolas Regnault.
\newblock Fractional {Chern} {Insulators} vs. {Non}-{Magnetic} {States} in {Twisted} {Bilayer} {MoTe}\$\_2\$, September 2023.
\newblock arXiv:2309.14429 [cond-mat].

\bibitem{wang_topological_2023}
Taige Wang, Trithep Devakul, Michael~P. Zaletel, and Liang Fu.
\newblock Topological magnets and magnons in twisted bilayer {MoTe}\$\_2\$ and {WSe}\$\_2\$, June 2023.
\newblock arXiv:2306.02501 [cond-mat].

\bibitem{goldman_zero-field_2023}
Hart Goldman, Aidan~P. Reddy, Nisarga Paul, and Liang Fu.
\newblock Zero-field composite {Fermi} liquid in twisted semiconductor bilayers.
\newblock {\em Phys. Rev. Lett.}, 131(13):136501, September 2023.
\newblock arXiv:2306.02513 [cond-mat].

\bibitem{reddy_toward_2023}
Aidan~P. Reddy and Liang Fu.
\newblock Toward a global phase diagram of the fractional quantum anomalous {Hall} effect, August 2023.
\newblock arXiv:2308.10406 [cond-mat].

\bibitem{dong_composite_2023}
Junkai Dong, Jie Wang, Patrick~J. Ledwith, Ashvin Vishwanath, and Daniel~E. Parker.
\newblock Composite {Fermi} {Liquid} at {Zero} {Magnetic} {Field} in {Twisted} $\textrm{MoTe}_2$.
\newblock {\em Phys. Rev. Lett.}, 131(13):136502, September 2023.
\newblock Publisher: American Physical Society.

\bibitem{morales-duran_magic_2023}
Nicol{\'a}s Morales-Dur{\'a}n, Nemin Wei, and Allan~H. MacDonald.
\newblock Magic {Angles} and {Fractional} {Chern} {Insulators} in {Twisted} {Homobilayer} {TMDs}, August 2023.
\newblock arXiv:2308.03143 [cond-mat].

\bibitem{hallcrystal2022}
Anna~M. Seiler, Fabian~R. Geisenhof, Felix Winterer, Kenji Watanabe, Takashi Taniguchi, Tianyi Xu, Fan Zhang, and R.~Thomas Weitz.
\newblock Quantum cascade of correlated phases in trigonally warped bilayer graphene.
\newblock {\em Nature}, 608(7922):298--302, August 2022.

\bibitem{Chen:2020aa}
Guorui Chen, Aaron~L. Sharpe, Eli~J. Fox, Ya-Hui Zhang, Shaoxin Wang, Lili Jiang, Bosai Lyu, Hongyuan Li, Kenji Watanabe, Takashi Taniguchi, Zhiwen Shi, T.~Senthil, David Goldhaber-Gordon, Yuanbo Zhang, and Feng Wang.
\newblock Tunable correlated chern insulator and ferromagnetism in a moir{\'e}superlattice.
\newblock {\em Nature}, 579(7797):56--61, 2020.

\bibitem{zhou2021half}
Haoxin Zhou, Tian Xie, Areg Ghazaryan, Tobias Holder, James~R Ehrets, Eric~M Spanton, Takashi Taniguchi, Kenji Watanabe, Erez Berg, Maksym Serbyn, et~al.
\newblock Half-and quarter-metals in rhombohedral trilayer graphene.
\newblock {\em Nature}, 598(7881):429--433, 2021.

\bibitem{han2023correlated}
Tonghang Han, Zhengguang Lu, Giovanni Scuri, Jiho Sung, Jue Wang, Tianyi Han, Kenji Watanabe, Takashi Taniguchi, Hongkun Park, and Long Ju.
\newblock Correlated insulator and chern insulators in pentalayer rhombohedral-stacked graphene.
\newblock {\em Nature Nanotechnology}, oct 2023.

\bibitem{han2023large}
Tonghang Han, Zhengguang Lu, Yuxuan Yao, Jixiang Yang, Junseok Seo, Chiho Yoon, Kenji Watanabe, Takashi Taniguchi, Liang Fu, Fan Zhang, and Long Ju.
\newblock Large quantum anomalous hall effect in spin-orbit proximitized rhombohedral graphene, 2023.

\bibitem{Chen2022ferro}
Guorui Chen, Aaron~L. Sharpe, Eli~J. Fox, Shaoxin Wang, Bosai Lyu, Lili Jiang, Hongyuan Li, Kenji Watanabe, Takashi Taniguchi, Michael~F. Crommie, Marc~A. Kastner, Zhiwen Shi, David Goldhaber-Gordon, Yuanbo Zhang, and Feng Wang.
\newblock Tunable orbital ferromagnetism at noninteger filling of a moiré superlattice.
\newblock {\em Nano Letters}, 22(1):238--245, 2022.
\newblock PMID: 34978444.

\bibitem{SpantonFCI}
Eric~M. Spanton, Alexander~A. Zibrov, Haoxin Zhou, Takashi Taniguchi, Kenji Watanabe, Michael~P. Zaletel, and Andrea~F. Young.
\newblock Observation of fractional chern insulators in a van der waals heterostructure.
\newblock {\em Science}, 360(6384):62--66, apr 2018.

\bibitem{Zhou2022BLG}
Haoxin Zhou, Ludwig Holleis, Yu~Saito, Liam Cohen, William Huynh, Caitlin~L. Patterson, Fangyuan Yang, Takashi Taniguchi, Kenji Watanabe, and Andrea~F. Young.
\newblock Isospin magnetism and spin-polarized superconductivity in bernal bilayer graphene.
\newblock {\em Science}, 375(6582):774--778, 2022.

\bibitem{Zhang2023BLG}
Yiran Zhang, Robert Polski, Alex Thomson, {\'E}tienne Lantagne-Hurtubise, Cyprian Lewandowski, Haoxin Zhou, Kenji Watanabe, Takashi Taniguchi, Jason Alicea, and Stevan Nadj-Perge.
\newblock Enhanced superconductivity in spin--orbit proximitized bilayer graphene.
\newblock {\em Nature}, 613(7943):268--273, Jan 2023.

\bibitem{Holleis2023BLG}
Ludwig Holleis, Caitlin~L. Patterson, Yiran Zhang, Heun~Mo Yoo, Haoxin Zhou, Takashi Taniguchi, Kenji Watanabe, Stevan Nadj-Perge, and Andrea~F. Young.
\newblock Ising superconductivity and nematicity in bernal bilayer graphene with strong spin orbit coupling, 2023.

\bibitem{Chen2019SC}
Guorui Chen, Aaron~L. Sharpe, Patrick Gallagher, Ilan~T. Rosen, Eli~J. Fox, Lili Jiang, Bosai Lyu, Hongyuan Li, Kenji Watanabe, Takashi Taniguchi, Jeil Jung, Zhiwen Shi, David Goldhaber-Gordon, Yuanbo Zhang, and Feng Wang.
\newblock Signatures of tunable superconductivity in a trilayer graphene moir{\'e} superlattice.
\newblock {\em Nature}, 572(7768):215--219, Aug 2019.

\bibitem{Lee2022TLG}
Yongjin Lee, Shi Che, Jairo Velasco~Jr., Xueshi Gao, Yanmeng Shi, David Tran, Jacopo Baima, Francesco Mauri, Matteo Calandra, Marc Bockrath, and Chun~Ning Lau.
\newblock Gate-tunable magnetism and giant magnetoresistance in suspended rhombohedral-stacked few-layer graphene.
\newblock {\em Nano Letters}, 22(13):5094--5099, Jul 2022.

\bibitem{Alexander20214LG}
Alexander Kerelsky, Carmen Rubio-Verdú, Lede Xian, Dante~M. Kennes, Dorri Halbertal, Nathan Finney, Larry Song, Simon Turkel, Lei Wang, Kenji Watanabe, and et~al.
\newblock Moiréless correlations in abca graphene.
\newblock {\em Proceedings of the National Academy of Sciences}, 118(4), 2021.

\bibitem{Liu20234LG}
Kai Liu, Jian Zheng, Yating Sha, Bosai Lyu, Fengping Li, Youngju Park, Yulu Ren, Kenji Watanabe, Takashi Taniguchi, Jinfeng Jia, Weidong Luo, Zhiwen Shi, Jeil Jung, and Guorui Chen.
\newblock Interaction-driven spontaneous broken-symmetry insulator and metals in abca tetralayer graphene, 2023.

\bibitem{Han2023multiferroicity}
Tonghang Han, Zhengguang Lu, Giovanni Scuri, Jiho Sung, Jue Wang, Tianyi Han, Kenji Watanabe, Takashi Taniguchi, Liang Fu, Hongkun Park, and Long Ju.
\newblock Orbital multiferroicity in pentalayer rhombohedral graphene.
\newblock {\em Nature}, 623(7985):41--47, Nov 2023.

\bibitem{sha2023observation}
Yating Sha, Jian Zheng, Kai Liu, Hong Du, Kenji Watanabe, Takashi Taniguchi, Jinfeng Jia, Zhiwen Shi, Ruidan Zhong, and Guorui Chen.
\newblock Observation of chern insulator in crystalline abca-tetralayer graphene with spin-orbit coupling.
\newblock {\em arXiv preprint arXiv:2310.17971}, 2023.

\bibitem{Senthil_NearlyFlatBand}
Ya-Hui Zhang, Dan Mao, Yuan Cao, Pablo Jarillo-Herrero, and T.~Senthil.
\newblock Nearly flat chern bands in moir\'e superlattices.
\newblock {\em Phys. Rev. B}, 99:075127, Feb 2019.

\bibitem{Chittari2019flat}
Bheema~Lingam Chittari, Guorui Chen, Yuanbo Zhang, Feng Wang, and Jeil Jung.
\newblock Gate-tunable topological flat bands in trilayer graphene boron-nitride moir\'e superlattices.
\newblock {\em Phys. Rev. Lett.}, 122:016401, Jan 2019.

\bibitem{Repellin2020narrow}
C\'ecile Repellin, Zhihuan Dong, Ya-Hui Zhang, and T.~Senthil.
\newblock Ferromagnetism in narrow bands of moir\'e superlattices.
\newblock {\em Phys. Rev. Lett.}, 124:187601, May 2020.

\bibitem{Zhang2019Hubbard}
Ya-Hui Zhang and T.~Senthil.
\newblock Bridging hubbard model physics and quantum hall physics in trilayer $\text{graphene}/h\ensuremath{-}\mathrm{BN}$ moir\'e superlattice.
\newblock {\em Phys. Rev. B}, 99:205150, May 2019.

\bibitem{moon2014electronic}
Pilkyung Moon and Mikito Koshino.
\newblock Electronic properties of graphene/hexagonal-boron-nitride moir{\'e} superlattice.
\newblock {\em Physical Review B}, 90(15):155406, 2014.

\bibitem{JungAbInitio}
Jeil Jung, Arnaud Raoux, Zhenhua Qiao, and A.~H. MacDonald.
\newblock Ab initio theory of moir\'e superlattice bands in layered two-dimensional materials.
\newblock {\em Phys. Rev. B}, 89:205414, May 2014.

\bibitem{jung2017moire}
Jeil Jung, Evan Laksono, Ashley~M DaSilva, Allan~H MacDonald, Marcin Mucha-Kruczy{\'n}ski, and Shaffique Adam.
\newblock Moir{\'e} band model and band gaps of graphene on hexagonal boron nitride.
\newblock {\em Physical Review B}, 96(8):085442, 2017.

\bibitem{krisna2023moire}
Lukas~PA Krisna and Mikito Koshino.
\newblock Moir{\'e} phonons in graphene/hexagonal boron nitride moir{\'e} superlattice.
\newblock {\em Physical Review B}, 107(11):115301, 2023.

\bibitem{lee2016ballistic}
Menyoung Lee, John~R. Wallbank, Patrick Gallagher, Kenji Watanabe, Takashi Taniguchi, Vladimir~I. Fal’ko, and David Goldhaber-Gordon.
\newblock Ballistic miniband conduction in a graphene superlattice.
\newblock {\em Science}, 353(6307):1526–1529, September 2016.

\bibitem{patri2023moire}
Adarsh~S. Patri and T.~Senthil.
\newblock Strong correlations in abc-stacked trilayer graphene: Moir\'e is important.
\newblock {\em Phys. Rev. B}, 107:165122, Apr 2023.

\bibitem{Cecile_PRL20}
C\'ecile Repellin, Zhihuan Dong, Ya-Hui Zhang, and T.~Senthil.
\newblock Ferromagnetism in narrow bands of moir\'e superlattices.
\newblock {\em Phys. Rev. Lett.}, 124:187601, May 2020.

\bibitem{PhysRevX.1.021014}
N.~Regnault and B.~Andrei Bernevig.
\newblock Fractional chern insulator.
\newblock {\em Phys. Rev. X}, 1:021014, Dec 2011.

\bibitem{Sheng:2011tr}
D.~N. Sheng, Zheng-Cheng Gu, Kai Sun, and L.~Sheng.
\newblock Fractional quantum hall effect in the absence of landau levels.
\newblock {\em Nature Communications}, 2(1):389, 2011.

\bibitem{NeupertFQHZeroField}
Titus Neupert, Luiz Santos, Claudio Chamon, and Christopher Mudry.
\newblock Fractional quantum hall states at zero magnetic field.
\newblock {\em Phys. Rev. Lett.}, 106:236804, Jun 2011.

\bibitem{zhao_review}
EMIL~J. BERGHOLTZ and ZHAO LIU.
\newblock Topological flat band models and fractional chern insulators.
\newblock {\em International Journal of Modern Physics B}, 27(24):1330017, 2013.

\bibitem{Neupertreview_2015}
Titus Neupert, Claudio Chamon, Thomas Iadecola, Luiz~H Santos, and Christopher Mudry.
\newblock Fractional (chern and topological) insulators.
\newblock {\em Physica Scripta}, T164:014005, aug 2015.

\bibitem{Liu_review_2023}
Zhao Liu and Emil~J. Bergholtz.
\newblock Recent developments in fractional chern insulators.
\newblock In {\em Reference Module in Materials Science and Materials Engineering}. Elsevier, 2023.

\bibitem{FCI_TBG_exp}
Yonglong Xie, Andrew~T. Pierce, Jeong~Min Park, Daniel~E. Parker, Eslam Khalaf, Patrick Ledwith, Yuan Cao, Seung~Hwan Lee, Shaowen Chen, Patrick~R. Forrester, Kenji Watanabe, Takashi Taniguchi, Ashvin Vishwanath, Pablo Jarillo-Herrero, and Amir Yacoby.
\newblock Fractional chern insulators in magic-angle twisted bilayer graphene.
\newblock {\em Nature}, 600(7889):439--443, 2021.

\bibitem{parker2021field}
Daniel Parker, Patrick Ledwith, Eslam Khalaf, Tomohiro Soejima, Johannes Hauschild, Yonglong Xie, Andrew Pierce, Michael~P Zaletel, Amir Yacoby, and Ashvin Vishwanath.
\newblock Field-tuned and zero-field fractional chern insulators in magic angle graphene.
\newblock {\em arXiv preprint arXiv:2112.13837}, 2021.

\bibitem{bultinck2020mechanism}
Nick Bultinck, Shubhayu Chatterjee, and Michael~P. Zaletel.
\newblock Mechanism for anomalous hall ferromagnetism in twisted bilayer graphene.
\newblock {\em Phys. Rev. Lett.}, 124:166601, Apr 2020.

\bibitem{xie2020nature}
Ming Xie and A.~H. MacDonald.
\newblock Nature of the correlated insulator states in twisted bilayer graphene.
\newblock {\em Phys. Rev. Lett.}, 124:097601, Mar 2020.

\bibitem{zou2018band}
Liujun Zou, Hoi~Chun Po, Ashvin Vishwanath, and T.~Senthil.
\newblock Band structure of twisted bilayer graphene: Emergent symmetries, commensurate approximants, and wannier obstructions.
\newblock {\em Phys. Rev. B}, 98:085435, Aug 2018.

\bibitem{zhang2019twisted}
Ya-Hui Zhang, Dan Mao, and T.~Senthil.
\newblock Twisted bilayer graphene aligned with hexagonal boron nitride: Anomalous hall effect and a lattice model.
\newblock {\em Phys. Rev. Res.}, 1:033126, Nov 2019.

\bibitem{wu2019topological}
Fengcheng Wu, Timothy Lovorn, Emanuel Tutuc, Ivar Martin, and AH~MacDonald.
\newblock Topological insulators in twisted transition metal dichalcogenide homobilayers.
\newblock {\em Physical review letters}, 122(8):086402, 2019.

\bibitem{wang2023topological}
Taige Wang, Trithep Devakul, Michael~P Zaletel, and Liang Fu.
\newblock Topological magnets and magnons in twisted bilayer mote $ \_2 $ and wse $ \_2$.
\newblock {\em arXiv preprint arXiv:2306.02501}, 2023.

\bibitem{hallcrystal1986PRL}
B.~I. Halperin, Z.~Tesanovi{\'{c}}, and F.~Axel.
\newblock Compatibility of crystalline order and the quantized hall effect.
\newblock {\em Phys. Rev. Lett.}, 57:922--922, Aug 1986.

\bibitem{hallcrystal1989PRB}
Zlatko Tesanovi{\'{c}}, Fran{\c{c}}oise Axel, and B.~I. Halperin.
\newblock ``hall crystal'' versus wigner crystal.
\newblock {\em Phys. Rev. B}, 39:8525--8551, Apr 1989.

\bibitem{zhang2010band}
Fan Zhang, Bhagawan Sahu, Hongki Min, and Allan~H MacDonald.
\newblock Band structure of a b c-stacked graphene trilayers.
\newblock {\em Physical Review B}, 82(3):035409, 2010.

\bibitem{jung2013gapped}
Jeil Jung and Allan~H MacDonald.
\newblock Gapped broken symmetry states in abc-stacked trilayer graphene.
\newblock {\em Physical Review B}, 88(7):075408, 2013.

\bibitem{footnoteSM}
{See supplemental material for a detailed description on the microscopic model, the Hartree-Fock calculations, and the many-body calculations, which includes Refs.~\cite{chen2019signatures,chatterjee2022inter,ghazaryan2023multilayer,wang2023electrical,bultinck2020ground,vafek2020renormalization, MPOCompression2, parker2021strain, Dan_parker21, wang2022kekul,vanderbilt_2018,MPOCompression1,tenpy,castro2009the,hwang2011direct,yang2021experimental,LaughlinDisorder, DavidNote}.}

\bibitem{Note1}
We estimate the conversion between $D$ and $u_D$ by comparing the reported HF phase diagram with the experiment in rhombohedral trilayer graphene~\cite {chatterjee2022inter,zhou2021half}.

\bibitem{park2023topological}
Youngju Park, Yeonju Kim, Bheema~Lingam Chittari, and Jeil Jung.
\newblock Topological flat bands in rhombohedral tetralayer and multilayer graphene on hexagonal boron nitride moire superlattices, 2023.

\bibitem{Note2}
Near $u_D \approx 0$ valence bands, which we ignore, also become relevant; our model is most reliable for the moir\'e-distant regime $u_D > 0$.

\bibitem{li2004entanglement}
Hui Li and F.~D.~M. Haldane.
\newblock Entanglement spectrum as a generalization of entanglement entropy: Identification of topological order in non-abelian fractional quantum hall effect states.
\newblock {\em Phys. Rev. Lett.}, 101:010504, Jul 2008.

\bibitem{PhysRevB.90.165139}
Rahul Roy.
\newblock Band geometry of fractional topological insulators.
\newblock {\em Phys. Rev. B}, 90:165139, Oct 2014.

\bibitem{Parameswaran_2013}
Siddharth~A. Parameswaran, Rahul Roy, and Shivaji~L. Sondhi.
\newblock Fractional quantum hall physics in topological flat bands.
\newblock {\em Comptes Rendus Physique}, 14(9-10):816--839, nov 2013.

\bibitem{Jackson:2015aa}
T.~S. Jackson, Gunnar M{\"o}ller, and Rahul Roy.
\newblock Geometric stability of topological lattice phases.
\newblock {\em Nature Communications}, 6(1):8629, 2015.

\bibitem{Martin_PositionMomentumDuality}
Martin Claassen, Ching~Hua Lee, Ronny Thomale, Xiao-Liang Qi, and Thomas~P. Devereaux.
\newblock Position-momentum duality and fractional quantum hall effect in chern insulators.
\newblock {\em Phys. Rev. Lett.}, 114:236802, Jun 2015.

\bibitem{Grisha_TBG}
Grigory Tarnopolsky, Alex~Jura Kruchkov, and Ashvin Vishwanath.
\newblock Origin of magic angles in twisted bilayer graphene.
\newblock {\em Phys. Rev. Lett.}, 122:106405, Mar 2019.

\bibitem{Grisha_TBG2}
Patrick~J. Ledwith, Grigory Tarnopolsky, Eslam Khalaf, and Ashvin Vishwanath.
\newblock Fractional chern insulator states in twisted bilayer graphene: An analytical approach.
\newblock {\em Phys. Rev. Research}, 2:023237, May 2020.

\bibitem{kahlerband1}
Tomoki Ozawa and Bruno Mera.
\newblock Relations between topology and the quantum metric for chern insulators.
\newblock {\em Phys. Rev. B}, 104:045103, Jul 2021.

\bibitem{kahlerband2}
Bruno Mera and Tomoki Ozawa.
\newblock K\"ahler geometry and chern insulators: Relations between topology and the quantum metric.
\newblock {\em Phys. Rev. B}, 104:045104, Jul 2021.

\bibitem{kahlerband3}
Bruno Mera and Tomoki Ozawa.
\newblock Engineering geometrically flat chern bands with fubini-study k\"ahler structure.
\newblock {\em Phys. Rev. B}, 104:115160, Sep 2021.

\bibitem{JieWang_exactlldescription}
Jie Wang, Jennifer Cano, Andrew~J. Millis, Zhao Liu, and Bo~Yang.
\newblock Exact landau level description of geometry and interaction in a flatband.
\newblock {\em Phys. Rev. Lett.}, 127:246403, Dec 2021.

\bibitem{Jie_hierarchyidealband}
Jie Wang and Zhao Liu.
\newblock Hierarchy of ideal flatbands in chiral twisted multilayer graphene models.
\newblock {\em Phys. Rev. Lett.}, 128:176403, Apr 2022.

\bibitem{LedwithVishwanathParker22}
Patrick~J. {Ledwith}, Ashvin {Vishwanath}, and Daniel~E. {Parker}.
\newblock {Vortexability: A Unifying Criterion for Ideal Fractional Chern Insulators}.
\newblock {\em arXiv e-prints}, page arXiv:2209.15023, September 2022.

\bibitem{JW_origin_22}
Jie Wang, Semyon Klevtsov, and Zhao Liu.
\newblock Origin of model fractional chern insulators in all topological ideal flatbands: Explicit color-entangled wave function and exact density algebra.
\newblock {\em Phys. Rev. Res.}, 5:023167, Jun 2023.

\bibitem{junkaidonghighC22}
Junkai Dong, Patrick~J. Ledwith, Eslam Khalaf, Jong~Yeon Lee, and Ashvin Vishwanath.
\newblock Many-body ground states from decomposition of ideal higher chern bands: Applications to chirally twisted graphene multilayers.
\newblock {\em Phys. Rev. Res.}, 5:023166, Jun 2023.

\bibitem{valentin23ideal}
B.~{Estienne}, N.~{Regnault}, and V.~{Cr{\'e}pel}.
\newblock {Ideal Chern bands are Landau levels in curved space}.
\newblock {\em arXiv e-prints}, page arXiv:2304.01251, April 2023.

\bibitem{andrews2023stability}
Bartholomew Andrews, Mathi Raja, Nimit Mishra, Michael Zaletel, and Rahul Roy.
\newblock Stability of fractional chern insulators with a non-landau level continuum limit.
\newblock {\em arXiv preprint arXiv:2310.05758}, 2023.

\bibitem{KaiSunNearlyFlat}
Kai Sun, Zhengcheng Gu, Hosho Katsura, and S.~Das~Sarma.
\newblock Nearly flatbands with nontrivial topology.
\newblock {\em Phys. Rev. Lett.}, 106:236803, Jun 2011.

\bibitem{WangFQHBoson}
Yi-Fei Wang, Zheng-Cheng Gu, Chang-De Gong, and D.~N. Sheng.
\newblock Fractional quantum hall effect of hard-core bosons in topological flat bands.
\newblock {\em Phys. Rev. Lett.}, 107:146803, Sep 2011.

\bibitem{TangHighTemperature}
Evelyn Tang, Jia-Wei Mei, and Xiao-Gang Wen.
\newblock High-temperature fractional quantum hall states.
\newblock {\em Phys. Rev. Lett.}, 106:236802, Jun 2011.

\bibitem{PhysRevB.41.7910}
J.~P. Eisenstein, H.~L. Stormer, L.~N. Pfeiffer, and K.~W. West.
\newblock Evidence for a spin transition in the \ensuremath{\nu}=2/3 fractional quantum hall effect.
\newblock {\em Phys. Rev. B}, 41:7910--7913, Apr 1990.

\bibitem{PhysRevLett.72.3405}
Y.~W. Suen, H.~C. Manoharan, X.~Ying, M.~B. Santos, and M.~Shayegan.
\newblock Origin of the \ensuremath{\nu}=1/2 fractional quantum hall state in wide single quantum wells.
\newblock {\em Phys. Rev. Lett.}, 72:3405--3408, May 1994.

\bibitem{PhysRevB.53.15845}
I.~A. McDonald and F.~D.~M. Haldane.
\newblock Topological phase transition in the \ensuremath{\nu}=2/3 quantum hall effect.
\newblock {\em Phys. Rev. B}, 53:15845--15855, Jun 1996.

\bibitem{Trail_2003}
J.~R. Trail, M.~D. Towler, and R.~J. Needs.
\newblock Unrestricted hartree-fock theory of wigner crystals.
\newblock {\em Physical Review B}, 68(4), jul 2003.

\bibitem{tanatar1989}
B.~Tanatar and D.~M. Ceperley.
\newblock Ground state of the two-dimensional electron gas.
\newblock {\em Phys. Rev. B}, 39:5005--5016, Mar 1989.

\bibitem{attaccalite2002}
Claudio Attaccalite, Saverio Moroni, Paola Gori-Giorgi, and Giovanni~B. Bachelet.
\newblock Correlation energy and spin polarization in the 2d electron gas.
\newblock {\em Phys. Rev. Lett.}, 88:256601, Jun 2002.

\bibitem{drummond2009}
N.~D. Drummond and R.~J. Needs.
\newblock Phase diagram of the low-density two-dimensional homogeneous electron gas.
\newblock {\em Phys. Rev. Lett.}, 102:126402, Mar 2009.

\bibitem{soejima2024anomaloushallcrystalsrhombohedral}
Tomohiro Soejima, Junkai Dong, Taige Wang, Tianle Wang, Michael~P. Zaletel, Ashvin Vishwanath, and Daniel~E. Parker.
\newblock Anomalous hall crystals in rhombohedral multilayer graphene ii: General mechanism and a minimal model, 2024.

\bibitem{dong2023theory}
Zhihuan Dong, Adarsh~S. Patri, and T.~Senthil.
\newblock Theory of fractional quantum anomalous hall phases in pentalayer rhombohedral graphene moir\'e structures, 2023.

\bibitem{zhou2023fractional}
Boran Zhou, Hui Yang, and Ya-Hui Zhang.
\newblock Fractional quantum anomalous hall effects in rhombohedral multilayer graphene in the moir\'eless limit and in coulomb imprinted superlattice, 2023.

\bibitem{guo_theory_2023}
Zhongqing Guo, Xin Lu, Bo~Xie, and Jianpeng Liu.
\newblock Theory of fractional {Chern} insulator states in pentalayer graphene moir{\textbackslash}'e superlattice, December 2023.
\newblock arXiv:2311.14368 [cond-mat].

\bibitem{kwan_moire_2023}
Yves~H. Kwan, Jiabin Yu, Jonah Herzog-Arbeitman, Dmitri~K. Efetov, Nicolas Regnault, and B.~Andrei Bernevig.
\newblock Moir{\textbackslash}'e {Fractional} {Chern} {Insulators} {III}: {Hartree}-{Fock} {Phase} {Diagram}, {Magic} {Angle} {Regime} for {Chern} {Insulator} {States}, the {Role} of the {Moir}{\textbackslash}'e {Potential} and {Goldstone} {Gaps} in {Rhombohedral} {Graphene} {Superlattices}, December 2023.
\newblock arXiv:2312.11617 [cond-mat].

\bibitem{chen2019signatures}
Guorui Chen, Aaron~L Sharpe, Patrick Gallagher, Ilan~T Rosen, Eli~J Fox, Lili Jiang, Bosai Lyu, Hongyuan Li, Kenji Watanabe, Takashi Taniguchi, et~al.
\newblock Signatures of tunable superconductivity in a trilayer graphene moir{\'e} superlattice.
\newblock {\em Nature}, 572(7768):215--219, 2019.

\bibitem{chatterjee2022inter}
Shubhayu Chatterjee, Taige Wang, Erez Berg, and Michael~P Zaletel.
\newblock Inter-valley coherent order and isospin fluctuation mediated superconductivity in rhombohedral trilayer graphene.
\newblock {\em Nature communications}, 13(1):6013, 2022.

\bibitem{ghazaryan2023multilayer}
Areg Ghazaryan, Tobias Holder, Erez Berg, and Maksym Serbyn.
\newblock Multilayer graphenes as a platform for interaction-driven physics and topological superconductivity.
\newblock {\em Physical Review B}, 107(10):104502, 2023.

\bibitem{wang2023electrical}
Taige Wang, Marc Vila, Michael~P Zaletel, and Shubhayu Chatterjee.
\newblock Electrical control of magnetism in spin-orbit coupled graphene multilayers.
\newblock {\em arXiv preprint arXiv:2303.04855}, 2023.

\bibitem{bultinck2020ground}
Nick Bultinck, Eslam Khalaf, Shang Liu, Shubhayu Chatterjee, Ashvin Vishwanath, and Michael~P Zaletel.
\newblock Ground state and hidden symmetry of magic-angle graphene at even integer filling.
\newblock {\em Physical Review X}, 10(3):031034, 2020.

\bibitem{vafek2020renormalization}
Oskar Vafek and Jian Kang.
\newblock Renormalization group study of hidden symmetry in twisted bilayer graphene with coulomb interactions.
\newblock {\em Physical Review Letters}, 125(25):257602, 2020.

\bibitem{LaughlinDisorder}
R.~B. Laughlin.
\newblock Quantized hall conductivity in two dimensions.
\newblock {\em Phys. Rev. B}, 23:5632--5633, May 1981.

\bibitem{DavidNote}
David Tong.
\newblock Lectures on the quantum hall effect.
\newblock {\em arXiv:1606.06687}, 2016.

\bibitem{castro2009the}
A.~H. Castro~Neto, F.~Guinea, N.~M.~R. Peres, K.~S. Novoselov, and A.~K. Geim.
\newblock The electronic properties of graphene.
\newblock {\em Rev. Mod. Phys.}, 81:109--162, Jan 2009.

\bibitem{hwang2011direct}
Choongyu Hwang, Cheol-Hwan Park, David~A. Siegel, Alexei~V. Fedorov, Steven~G. Louie, and Alessandra Lanzara.
\newblock Direct measurement of quantum phases in graphene via photoemission spectroscopy.
\newblock {\em Phys. Rev. B}, 84:125422, Sep 2011.

\bibitem{yang2021experimental}
Fangyuan Yang, Alexander~A. Zibrov, Ruiheng Bai, Takashi Taniguchi, Kenji Watanabe, Michael~P. Zaletel, and Andrea~F. Young.
\newblock Experimental determination of the energy per particle in partially filled landau levels.
\newblock {\em Phys. Rev. Lett.}, 126:156802, Apr 2021.

\bibitem{MPOCompression2}
Tomohiro Soejima, Daniel~E. Parker, Nick Bultinck, Johannes Hauschild, and Michael~P. Zaletel.
\newblock Efficient simulation of moir\'e materials using the density matrix renormalization group.
\newblock {\em Phys. Rev. B}, 102:205111, Nov 2020.

\bibitem{parker2021strain}
Daniel~E. Parker, Tomohiro Soejima, Johannes Hauschild, Michael~P. Zaletel, and Nick Bultinck.
\newblock Strain-induced quantum phase transitions in magic-angle graphene.
\newblock {\em Phys. Rev. Lett.}, 127:027601, Jul 2021.

\bibitem{Dan_parker21}
Daniel {Parker}, Patrick {Ledwith}, Eslam {Khalaf}, Tomohiro {Soejima}, Johannes {Hauschild}, Yonglong {Xie}, Andrew {Pierce}, Michael~P. {Zaletel}, Amir {Yacoby}, and Ashvin {Vishwanath}.
\newblock {Field-tuned and zero-field fractional Chern insulators in magic angle graphene}.
\newblock {\em arXiv e-prints}, page arXiv:2112.13837, December 2021.

\bibitem{wang2022kekul}
Tianle Wang, Daniel~E Parker, Tomohiro Soejima, Johannes Hauschild, Sajant Anand, Nick Bultinck, and Michael~P Zaletel.
\newblock Kekul$\backslash$'e spiral order in magic-angle graphene: a density matrix renormalization group study.
\newblock {\em arXiv preprint arXiv:2211.02693}, 2022.

\bibitem{vanderbilt_2018}
David Vanderbilt.
\newblock {\em Berry Phases in Electronic Structure Theory: Electric Polarization, Orbital Magnetization and Topological Insulators}.
\newblock Cambridge University Press, 2018.

\bibitem{MPOCompression1}
Daniel~E. Parker, Xiangyu Cao, and Michael~P. Zaletel.
\newblock Local matrix product operators: Canonical form, compression, and control theory.
\newblock {\em Phys. Rev. B}, 102:035147, Jul 2020.

\bibitem{tenpy}
Johannes Hauschild and Frank Pollmann.
\newblock Efficient numerical simulations with tensor networks: Tensor network python (tenpy).
\newblock {\em SciPost Physics Lecture Notes}, page 005, 2018.

\end{thebibliography}

\appendix

\section{Details of the Microscopic Model}
\label{app:microscopic_model}

This appendix fully specifies the models discussed in the main text.

\subsection{Rhombohedral Graphene}
Let $a \approx \SI{2.46}{\angstrom}$ be the lattice constant of graphene and consider real space and reciprocal lattices spanned by
\begin{align}
	\v{R}_1 = \left( a,0 \right), \qquad
	&\v{R}_2 = \left( \frac{1}{2}, \frac{\sqrt{3}}{2} \right),\\
	\v{G}_1 = \frac{2\pi}{a}\left(1, -\frac{1}{\sqrt{3}}\right), \qquad
	&\v{G}_2 = \frac{2\pi}{a}\left(0, \frac{2}{\sqrt{3}}\right).
\end{align}
For $N$-layer Rhombohedral graphene~\cite{zhang2010band,jung2013gapped} (see also:~\cite{chen2019signatures,zhou2021half,chatterjee2022inter,ghazaryan2023multilayer,park2023topological} and references therein), consider $p_z$ orbitals of carbon at positions
\begin{equation}
	\v{r}_{A,\ell} = \left( 0,\frac{\ell-1}{\sqrt{3}}a, c \right),  
	\v{r}_{B,\ell} = \left( 0,\frac{\ell}{\sqrt{3}}a, c \right)
\end{equation}
for layer $\ell \in [1,N_L]$ with inter-layer spacing $c$. Let the nearest neighbor vectors be $\v{\delta}_{n} = R_{n 2\pi/3} (0,\frac{1}{\sqrt{3}}a)^T$ for $n=0,1,2$ where $R_{\theta}$ is the counterclockwise rotation matrix by angle $\theta$. Below we index sublattice by $\sigma \in \st{A,B} = \st{-1,1}$. We define standard high-symmetry points $\v{K} = \tfrac{2}{3} \v{G}_1 + \tfrac{1}{3} \v{G}_2$ and $\v{K}' = \tfrac{1}{3} \v{G}_1 + \tfrac{2}{3} \v{G}_2$.

For each orbital, we have a corresponding second-quantized Fermion operator $\hat{c}^\dagger_{\v{R}, \sigma,\ell} = \hat{c}^\dagger(\v{R}+\v{r}_{\sigma,\ell})$. Its Fourier transform is $\hat{c}^\dagger_{\v{k}\sigma \ell} = N^{-\frac{1}{2}} \sum_{\v{R}} e^{+i\v{k} \cdot (\v{R}+\v{r}_{\sigma,j})} \hat{c}^\dagger(\v{R}+\v{r}_{\sigma,\ell})$, with normalization so that it obeys canonical commutation relations $\st{\hat{c}_{\v{k} \sigma \ell},\hat{c}_{\v{k} \sigma' \ell'}} = \delta_{\sigma\sigma'} \delta_{\ell\ell'}$. The Hamiltonian for $N$-layer rhombohedral graphene is $\hat{h}_{RG}^{(N_L)}~=~\sum_{\v{k} \in \bz} \hat{c}_{\v{k} \sigma j}^\dagger \left[ h_{RG}^{(N_L)}(\v{k}) \right]_{\sigma j, \sigma'j'} \hat{c}_{\v{k} \sigma' j'}$ where
\begin{equation}
	h_{RG}^{(N_L)}(\v{k}) =
	\begin{bmatrix} 
		h^{(0)}_{\ell} & h^{(1)} & h^{(2)}\\[0.5em]
		h^{(1)\dagger} & h^{(0)}_{\ell} & h^{(1)} & h^{(2)}\\[0.5em]
		h^{(2)\dagger} & h^{(1)\dagger} & h^{(0)}_{\ell} & \ddots & \ddots\\[0.5em]
				& h^{(2)\dagger} & \ddots & \ddots & h^{(1)} & h^{(2)}\\[0.5em]
				& & \ddots & h^{(1)^\dagger} & h^{(0)}_{\ell} & h^{(1)}\\[0.5em]
				& & & h^{(2)\dagger} & h^{(1)\dagger} & h^{(0)}_{\ell}\\
	\end{bmatrix}
	\label{eq:rhombohedral_graphene}
\end{equation}
which is a matrix in layer space with entries in sublattice space
\begin{subequations}
	\begin{align}
		h^{(0)}_{\ell} (\v{k}) &= \begin{pmatrix} u_{A\ell}  & -t_0 f_{\v{k}}\\ -t_0 \overline{f}_{\v{k}} & u_{B\ell} \end{pmatrix}\\[1.5em] 
		h^{(1)}(\v{k}) &= \begin{pmatrix} t_4 f_{\v{k}} & t_3 \overline{f_{\v{k}}} \\ t_1 & t_4 f_{\v{k}} \end{pmatrix}\\[1.5em]
		h^{(2)}(\v{k}) &= \begin{pmatrix} 0 & \frac{t_2}{2}\\ 0 & 0 \end{pmatrix}\\[1em]
		f_{\v{k}} &= \sum_{i=0}^2 e^{i \v{k}\cdot \v{\delta}_i}.
	\end{align}
\end{subequations}
We take graphene hopping $t_0 = \SI{3.1}{eV}$ and vertical interlayer hopping $t_1 = \SI{380}{meV}$. Following e.g.~\cite{wang2023electrical} we take higher hopping parameters $(t_2,t_3,t_4) = (-21,290,141)$ \si{meV}. Finally, $u_{\sigma\ell}$ are on-site potentials that reflect internal and external displacement fields. Although their value varies with layer~\cite{ghazaryan2023multilayer} to partially screen applied electric fields, for simplicity we consider fields with a uniform potential difference per layer $u_D$:
\begin{equation}
	u_{\sigma, \ell} = u_D \left( \ell + 1 - \frac{N_L-1}{2} \right).
\end{equation}

Aside from lattice translation, the model enjoys the following symmetries
\begin{subequations}
\begin{align}
	&\overline{h(\v{k})} = h(-\v{k}) \hspace{1em} (\text{Time reversal})\\
	&h(\v{k}) = h(C_{3z}\v{k})  \hspace{1em} (\text{Three fold rotation}) \\
	&M_{C_{2x}} h(\v{k}) M_{C_{2x}}^{-1} = h(C_{2x}\v{k})  \hspace{1em} (\text{Two-fold $x$-rotation})\\
	&h(\v{k}) = h(M_x \v{k}) \hspace{1em} (\text{Mirror})\\
	&V_{\v{G}} h(\v{k}) V_{\v{G}}^\dagger = h(\v{k}+\v{G}) \hspace{1em} (\text{$\v{G}$-translation})
\end{align}
\end{subequations}
where time-reversal acts anti-unitarily, $C_{2x}$ reverses layers as $[M_{C_{2x}}]_{\sigma \ell, \sigma'\ell'} = \delta_{\sigma,-\sigma'} \delta_{\ell,N-(\ell'+1)}$, and other symmetries have trivial orbital part. We note that the last equation is not a symmetry per se, but simply a constraint on the Hamiltonian with $[V_{\v{G}}]_{\sigma' \ell', \sigma \ell} = \delta_{\sigma\sigma'} \delta_{\ell\ell'} e^{i \v{G} \cdot \v{r}_{\sigma,\ell}}$, reflecting the unit cell embedding.

\subsection{Moir\'e Rhombohedral Graphene Hamiltonian}

The direct lattice of the hBN substrate is is generated by
\begin{equation}
	\v{R}_j' = M R_{\theta} \v{R}_j; \quad M = (1+\varepsilon) I
\end{equation}
where $R_{\theta}$ is counterclockwise rotation by $\theta$ and $\varepsilon = a_{\textrm{hBN}}/a_{\mathrm{Gr}} \approx 1.018$. We take $\v{G}_j'$ such that $\v{R}_j' \cdot \v{G}_k' = 2\pi \delta_{jk}$ as usual. 

\begin{figure}[h]
	\centering
	\includegraphics[width=0.99\linewidth]{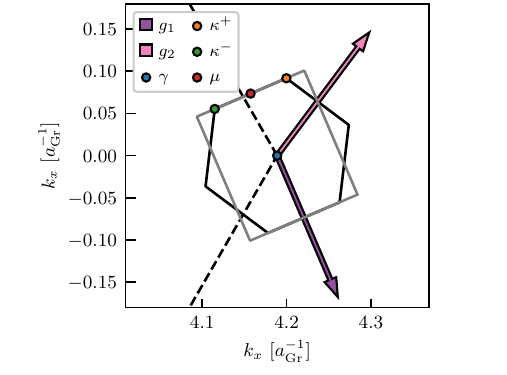}
    \caption{Geometry of the moir\'e Brillouin zone at $\theta=0.77^\circ$. The dashed line is the Brillouin zone of (unrotated) graphene, and the gray rectangle is the Brillouin zone used for DMRG (see text).}
    \label{fig:mbz_geometry}
\end{figure}

Define moir\'e reciprocal lattice vectors
\begin{equation}
	\v{g}_j = \v{G}_j - \v{G}_j' = [I - M^{-1} R_{\theta}] \v{G}_j,
\end{equation}
with corresponding direct vectors $\v{a}_j \cdot \v{g}_k = 2\pi \delta_{jk}$ whose moir\'e superlattice period is
\begin{equation}
	L_M = \n{\v{a}_j} = a \frac{1+\varepsilon}{\sqrt{\varepsilon^2 + 2(1+\varepsilon)(1-\cos \theta})}
\end{equation}
and where the lattice scale mismatch rotates the moire-Brillouin zone so that $\v{a}_1$ is rotated clockwise from the $x$ axis by an angle
\begin{equation}
	\phi = \arctan \left( \frac{-\sin \theta}{1 + \varepsilon - \cos \varepsilon} \right). 
    \label{eq:moire_phi}
\end{equation}
As $\theta$ changes from $0$ to $1$, the superlattice scale changes only gradually, but $\phi$ rotates quickly, as shown in Fig. \ref{fig:mbz_geometry}.

To conveniently express $C_3$-symmetric quantities, further define $\v{g}_3 = -\v{g}_1 - \v{g}_2$ and likewise for $\v{G}_3, \v{G}_3'$. We then have high-symmetry points at 
\begin{equation}  
\begin{aligned}
	\v{\gamma} &= \v{K}_{\mathrm{Gr}}, \quad
	\v{\kappa}^+ = \v{\gamma} -\frac{1}{3}\v{g}_1 + \frac{1}{3}\v{g}_2,\\
	\v{\kappa}^+ &= \v{\gamma} -\frac{2}{3}\v{g}_1 - \frac{1}{3}\v{g}_2, \quad
	\v{\mu} = \v{\gamma} - \frac{1}{2}\v{g}_1.
\end{aligned}
\end{equation}
Crucially, we have taken the center of the moir\'e Brillouin zone $\v{\gamma}$ to be at the $\v{K}$-point of graphene.

We may now express the moir\'e potential generated by the hBN substrate. We assume that the moir\'e potential of the hBN substrate only affects the closest layer. We follow~\cite{moon2014electronic} (which uses an opposite valley convention), we have
\begin{equation}
    V_{\textrm{hBN}}(\v{r}) = V_0 \sigma^0 + V_1 \v{f}(\v{r},\psi)\cdot\v{\sigma},
\end{equation}
where
\begin{equation}  
\begin{aligned}
    \v{f}(\v{r},\psi)\cdot\v{\sigma} = e^{-i\psi}\Big[&e^{i\v{g}_1\cdot\v r}\begin{pmatrix}
        1 & 1\\
        \omega &\omega
    \end{pmatrix} + e^{i\v{g}_2\cdot\v r}\begin{pmatrix}
        1 & \omega^2\\
        \omega^2 &\omega
    \end{pmatrix}\\ &+ e^{-i(\v{g}_1+\v{g}_2)\cdot\v r}\begin{pmatrix}
        1 & \omega\\
        1 &\omega
    \end{pmatrix}\Big]+\textrm{H.c.}
\end{aligned}
\end{equation}
Unless otherwise stated we take values $(V_0,V_1,\psi)=(\SI{28.9}{\milli\electronvolt}, \SI{21}{\milli\electronvolt},-0.29)$ following~\cite{moon2014electronic}. The Hamiltonian in the $K'$ valley is defined via time-reversal symmetry. 

\subsection{Many-Body Hamiltonian}

Let the single-particle eigenstates be given by
\begin{equation}
    \hat{h}_{\textit{kin}} \ket{\psi_{\v{k} \tau s b}} = \epsilon_{\v{k} \tau s b} \ket{\psi_{\v{k}\tau s b}}
\end{equation}
where $\tau \in \st{K,K'} = \st{-1,1}$ labels valley, $s\in \st{\uparrow, \downarrow} = \st{-1,1}$ labels spin, and $b$ labels bands. 
Consider creation operators
\begin{equation}
    c^\dagger_{\v{k}\tau s b}\ket{0}=\ket{\psi_{\v k\tau s b}},\quad 
    \st{c_{\v{k}\tau s b}, c_{\v{k}'\tau' s' b'}} = \delta_{\v{k}\v{k}'} \delta_{\tau \tau'}\delta_{ss'} \delta_{bb'}.
\end{equation}
We consider the interacting model
\begin{equation}
    \hat{H} = \hat{h}_{\textrm{kin}}+\frac{1}{2A} \sum_{\v{q}}U_{|\v q|} :\hat{\rho}_{\v q}\hat{\rho}_{-\v q}:, U_{q} = \frac{2\pi \tanh(qd)}{\epsilon_r\epsilon_0 q}
    \label{eq:app_many_body_Hamiltonian}
\end{equation}
where $U_{q}$ represents Coulomb interactions screened by both top and bottom gates at a distance of $d=\SI{250}{\angstrom}$.

To make the Hilbert space tractable, we restrict to a limited number of bands $N_b$ above charge neutrality. (Alternatively one could use a plane-wave basis and use some number of shells of Brillouin zones around the first.) More explicitly, we work with the projected density operator,
\begin{equation}
    \hat{\rho}_{\v{q}} = \sum_{\v{k}} \sum_{\alpha,\beta} \sum_{b,b'<N_b} \hat{c}_{\v{k} \alpha b}^\dagger \braket{\psi_{\v{k}\alpha b}|e^{-i\v{q}\cdot\v{r}} |\psi_{\v{k}+\v{q},\beta,b'}} \hat{c}_{\v{k}+\v{q}, \beta,b'}
\end{equation}
where $\alpha,\beta$ index valley, spin, and $b,b'$ labels band. In some contexts such as twisted bilayer graphene, the operation of restricting to low-energy bands is fraught with difficulties due to double-counting of interaction effects, necessitating procedures such as ``Hartree-Fock subtraction"~\cite{bultinck2020ground, parker2021field} or more sophisticated renormalization-group treatments~\cite{vafek2020renormalization}. Happily, here these issues are ameliorated in the limit where the displacement field opens a gap comparable to or larger than the interaction strength. In that case, the insulating state at charge neutrality is a natural choice of vacuum. The operation of integrating out the ``remote" higher-energy conduction bands $N_b+1, N_b + 2, \dots$, and valence bands at mean-field level is equivalent to simply \textit{restricting} the band indices in all sums in Eq. \eqref{eq:app_many_body_Hamiltonian}. We note, however, that beyond-mean-field effects like RPA-level screening from remote bands could substantially influence the system. This is often captured at a phenomenological level by increasing the relative dielectric constant $\epsilon_r$. 

\begin{figure*}
    \centering
    \includegraphics[width=\textwidth]{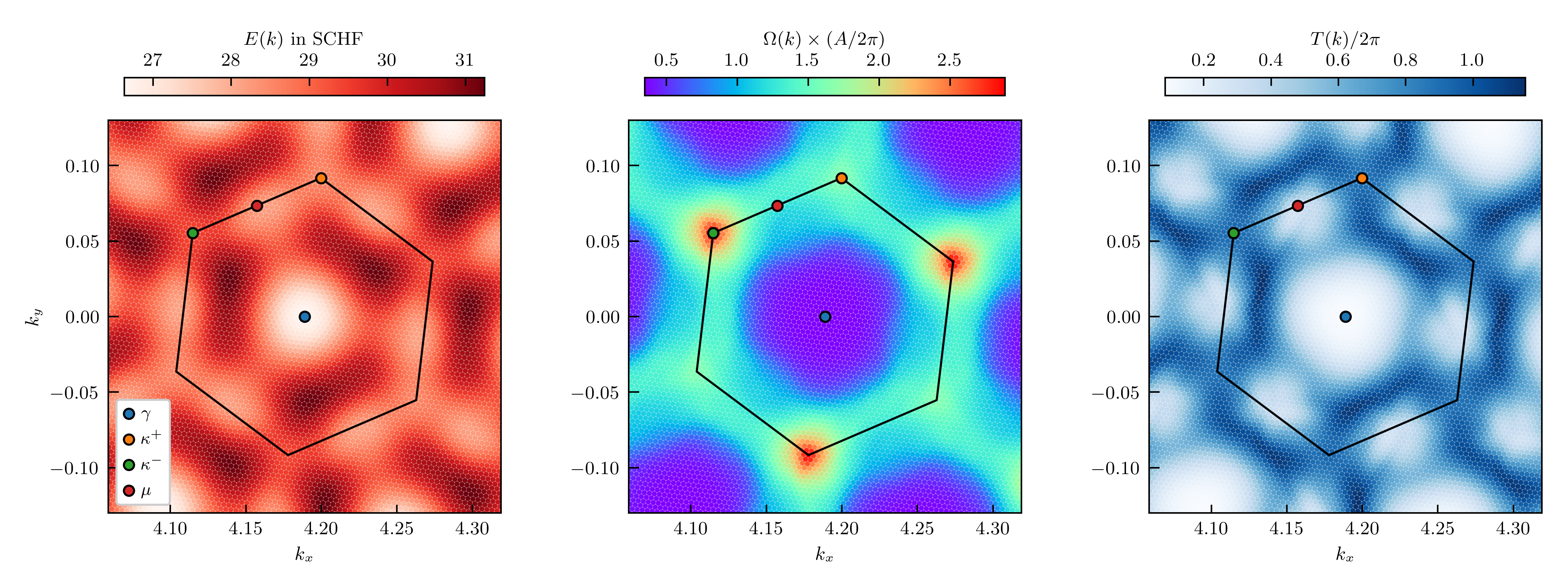}
    \caption{Details of the SCHF state at $\nu=1$. (Left) Dispersion. (Middle) Berry curvature. (Right) trace condition violation.   
    Parameters: $(u_D, \theta)=(\SI{50}{meV}, 0.77^\circ)$, $48\times 48$ unit cells.}
    \label{fig:app_nu_1_AHC_details}
\end{figure*}

\section{Details of the Hartree-Fock calculation}
\label{app:HartreeFock}
We use standard self-consistent Hartree-Fock calculations to obtain the SCHF ground state of the electronic system. 

\subsection{Numerical Setup}
\label{app:numerical_setup}

As usual, we consider single-particle density matrices $P(\v{k})_{\alpha\beta}=\braket{c^\dagger_{\v k \beta}c_{\v k \alpha}}$, where $\alpha$ and $\beta$ are collective indices for valley, spin, and band. These density matrices are in one-to-one correspondence with Slater determinant states. We define the Hartree and Fock Hamiltonians as 

\begin{subequations}
\begin{align}
    h_H[P](\v k) &=\frac{1}{A}\sum_{\v g}V_{\v g}\Lambda_{\v g}(\v k)\left(\sum_{\v k}\Tr[P(\v k)\Lambda_{\v g}(\v k)^\dagger]\right)\\
    h_F[P](\v k) &=-\frac{1}{A}\sum_{\v q}V_{\v q}\Lambda_{\v q}(\v k) P([\v k+\v q])\Lambda_{\v q}(\v k)^\dagger
\end{align}
\end{subequations}
where $[\Lambda_{\v q}(\v k)]_{\alpha\beta}=\braket{\psi_{\v k \alpha}|e^{-i\v{q}\cdot\v{r}}|\psi_{\v{k}+\v{q} \beta}}$ are form factors, and are treated as matrices whose labels are identical to the single-particle ones. The sum over $\v{g}$ runs over reciprocal vectors while $\v{q}$ runs over all momentum transfers. Via Wick's theorem, the energy of the Slater-determinant state is
\begin{equation}
    E[P]=\frac{1}{2}\Tr[P(2h_{\textrm{kin}}+h_{H}[P]+h_F[P])]
\end{equation}
where the trace is over momentum and all other band labels. 

To differentiate different ground states obtained in HF, we compute the Chern number and the HF gap. We regard any states with gap $\Delta_{\mathrm{HF}} < \SI{5}{meV}$ as a metal. For insulators, we show their Chern number in the phase diagram. We note that AHC and WLI have identical symmetry and therefore cannot be probed individually within HF by fixing symmetry. Instead, we use $50$ random seeds and a few physically motivated seeds to ensure a global minimum is found and the phase boundary between AHC and WLI is better resolved.

\subsection{Hartree-Fock Convergence}

We comment briefly on the stringent criteria we have used to ensure convergence of our SCHF numerics.

We use the optimal-damping algorithm (ODA) to converge to states satisfying the self-consistency condition 
\begin{equation}
    [P,h_\textrm{kin} + h_{H}[P]+h_{F}[P]]=0
\end{equation}
to tolerances approaching the square root of machine precision (i.e. machine precision in $E[P]$). We use both rectangular and Monkhorst-Pack grids with $24\times 24$, $30\times 30$, $36\times 36$, or $48\times 48$ unit cells, depending on the sensitivity of the state to finite-size effects. We impose $C_3$ symmetry explicitly in the Monkhorst-Pack case, but find comparable results in both cases for sufficiently large system sizes. We ensure that the range of momentum transfers $\v{q}$ considered is sufficiently large to converge the energy of the state, out to several Brillouin zone distances. To avoid non-global minima, we initialize SCHF in a variety of states for each parameter point, often using a mix of physical ansatzes as well as random initial states. We also assume electrons are spin- and valley-polarized.

\begin{figure}
    \includegraphics[width=0.5\textwidth]{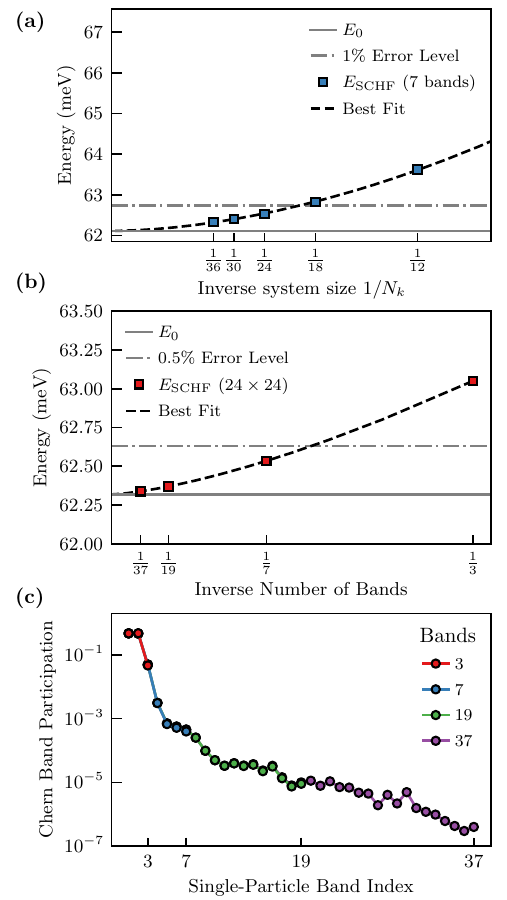}
    \caption{Convergence details of SCHF computations. (a) SCHF Ground state energy convergence as a function of system size $N_k \times N_k$ with $N_b = 7$ bands. 
    (b) SCHF ground state energy convergence as a function of the number of bands.
    (c) Participation of the single-particle bands in the isolated Chern band found within SCHF. Note that only the first three bands contribute significantly, and contributions are negligible after $N_b=7$.
    Parameters: $(u_D, \theta)=(\SI{50}{meV}, 0.77^\circ)$ unless specified.
    }
    \label{fig:app_SCHF_convergence}
\end{figure}

We now detail the convergence of our SCHF calculations as a function of the number of $k$-points and the number of single-particle bands. Fig.~\ref{fig:app_SCHF_convergence}(a) shows the ground state energy per electron within SCHF on a system of $N_k \times N_k$ unit cells with a Monkhorst-Pack lattice with the number of bands $N_b$=7. One can see that the energy is decreasing monotonically with the system size, and $1\%$ relative accuracy is achieved by $24\times 24$ unit cells. Furthermore, we note that the Berry curvature and energetic details of the isolated Chern band are qualitatively stable with system size  (with the notable exception of $18\times 18$ which is often qualitatively different, yielding different phase boundaries). We therefore conclude that $24\times 24$ unit cells is sufficient to capture quantitative details of phase boundaries to reasonably good fidelity.

Our SCHF is carried out in a Hilbert space spanned by the projector into the lowest $N_b$ conduction bands. We take $N_b = 7$ bands throughout the main text, a choice we now justify. At an initial qualitative level, we note that the conduction band wavefunctions come from folding the bands of rhombohedral graphene into the moir\'e Brillouin zone, then mixing them between $\v{k}$ and $\v{k}+\v{g}$ under the moir\'e potential. We note that the first seven bands are largely supported on the first moir\'e Brillouin zone together with the first shell of six higher Brillouin zones. Indeed, the first seven bands span a range of kinetic energy much greater than the interaction scale. This suggests the first 7 bands span an adequately large Hilbert space to understand the physics at filling $\nu =1$. We now investigate this claim quantitatively. Fig.~\ref{fig:app_SCHF_convergence}(b) show the SCHF energy as a function of $N_b$ (with values chosen to match centered hexagon numbers, i.e. complete shells of Brillouin zones). One can see that the energy is converged to better than $0.5\%$ relative error by $N_b = 7$. Furthermore, we investigate the participation of the higher bands in the Chern band that appears in SCHF at $\nu=1$. Explicitly, we compute the participation $P_n = \frac{1}{\mathcal{N}} \left( \sum_{\v{k}} |\braket{\phi_{\v{k},1}^{\mathrm{HF}}|\psi_{\v{k},n}}|^2 \right)$ of the single-particle bands in the wavefunction of the isolated Chern band $\phi_{\v{k},1}^{\mathrm{HF}}$ at $\nu=1$ with normalization so that $\sum_n |P_n|^2 =1$. We find these overlaps decrease precipitously after $n = 3$ and, furthermore, we find the participations are essentially stable after $N_b=7$. In conclusion, the SCHF ground state of our model is captured to good quantitative accuracy by $24 \times 24$ systems with $N_b=7$. 

As a final note, we emphasize that the errors from experimental and modelling uncertainties are likely to be vastly larger than numerical errors in this work.

\subsection{Metal versus Anomalous Hall Crystal Competition}

To examine the competition between the metallic ``quarter metal" state and the anomalous Hall crystal shown in Fig.~\ref{fig:intro}(b), we must determine the ground state energy both with and without continuous translation symmetry. To do this, we first compute the SCHF ground state of standalone rhombohedral graphene in a patch near the $K$ point that is much bigger than the moir\'e Brillouin zone, finding a metallic state with enforced continuous translation symmetry. We then compute another SCHF ground state with a folded Brillouin zone, which gives the possibility of breaking continuous translation symmetry. No hBN moir\'{e} potential is added. At small $u_D$, the energies from both methods agree to extremely high precision and indeed describe the same quarter-metal state. At a critical $u_D$, however, the translation-breaking ansatz finds a ground state with an interaction induced gap that is lower energy than the metal --- the anomalous Hall crystal described in the main text.

\subsection{Isospin symmetry breaking}

\begin{figure}
    \centering
    \includegraphics[width=0.99\linewidth]{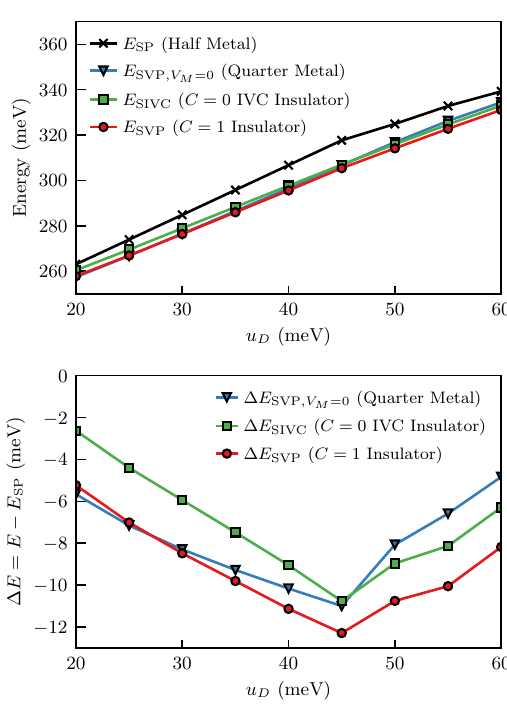}
    \caption{Energy per electron of states with different flavor ordering patterns. (top panel) The energy per electron of four competing ground state candidate states. (bottom panel) The energy per electron of three competing candidate states relative to the $N_f = 2$ spin polarized state.}
    \label{fig:spinvalley}
\end{figure}

A universal feature of RMG at high displacement field $D$ and low filling $\nu$ is isospin symmetry breaking, regardless of the moir\'{e} potential. A plethora of evidence, including quantum oscillation experiments~\cite{zhou2021half,Zhou2022BLG,Zhang2023BLG,Han2023multiferroicity} and SCHF numerics~\cite{chatterjee2022inter,wang2023electrical}, support a fully spin and valley polarized state around filling $n \sim \SI{e11}{cm^{-2}}$ (unity moir\'{e} filling at twist angle $\theta < 1^{\circ}$). Therefore, we assume full spin and valley polarization in the SCHF calculation presented in the main text. In this section, we justify this choice by performing SCHF with both spin and valley degrees of freedom in the parameter regime where we have observed AHC.

\begin{table}[!htbp]
    \renewcommand*{\arraystretch}{1.3}
    \centering
    \begin{tabular}{ccccc}
    \hline \hline
    \multirow{2}{*}{Group} & \multirow{2}{*}{flavor order} & \multicolumn{3}{c}{Symmetry}                        \\ \cline{3-5} 
                           &                                 & $SU(2)_s$    & $\tilde{\mathcal{T}}$ & $U(1)_v$     \\ \hline
    \multirow{3}{*}{(i)}   & SP                              & $\times$     & $\checkmark$          & $\checkmark$ \\
                           & VP                              & $\checkmark$ & $\times$              & $\checkmark$ \\
                           & SVL                             & $\times$     & $\times$              & $\checkmark$ \\ \hline
    (ii)                  & SVP                             & $\times$     & $\times$              & $\checkmark$ \\ \hline
    (iii)                   & SIVC                            & $\times$     & $\checkmark$          & $\times$     \\ \hline \hline
    \end{tabular}
    \caption{Symmetry of various magnetic orders, including the spin $SU(2)_s$ symmetry, spinless time-reversal $\tilde{\mathcal{T}}$ symmetry, and valley $U(1)_v$ symmetry.}
    \label{tab:symmetry}
\end{table}

If we neglect charge orders for the moment, different flavor orders can be classified by the isospin symmetry breaking pattern (Table~\ref{tab:symmetry}). We consider states in the following groups:
(i) an $N_F = 2$ state, where two out of four possible spin-valley flavors are occupied evenly. This includes a spin polarized (SP) state, a valley polarized (VP) state, and a spin-valley locked (SVL) state. These are candidate states for experimentally observed ``half metal'' states.
(ii) a spin-valley polarized (SVP) $N_F = 1$ state, where only one spin-valley species is occupied,  and (iii) a spin polarized IVC $N_F = 1$ state (SIVC) that breaks $\mathrm{U}(1)_v$. The latter two groups include the ``quarter metal'' states seen in experiment.

Since the aforementioned states have different symmetries/quantum numbers~Table~\ref{tab:symmetry}, we can enforce symmetries in the HF calculation and compare the energy among these states. On top of the magnetic order, we also allow translation symmetry breaking at moir\'{e} lattice vectors. 
Our Hamiltonian does not include any spin-orbit coupling or Hund's term, so the system has an enlarged $\mathrm{SU}(2)_{K} \times \mathrm{SU}(2)_{K'} $ symmetry. At HF level, candidate ground states within the group (i) are degenerate, whether it be SP, VP, or SVL. We therefore choose an SP state to represent the group (i).

In the top panel of Fig.~\ref{fig:spinvalley}, we show the energy per electron of $N_f=2$ SP states, $N_f = 1$ SIVC states, $N_f = 2$ SVP states. For comparison, we also show SVP metallic states that do not break translation symmetry computed with $V_M = 0$.

At $\nu = 1$, we always find the $N_F = 2$ SP state to have the highest energy, so we plot the energy of the other states relative to the SP state in the bottom panel of Fig.~\ref{fig:spinvalley}. We note that the SP states are always metallic since they correspond to $\nu=0.5$ per flavor. We find that the isospin symmetry breaking happens at an energy scale at least $5$ times bigger than the formation of an insulating state within the SVP phase (see also Fig.~\ref{fig:intro}(b)).

In terms of flavor order, The closest competing state of the SVP AHC is an SIVC AHC.
We find that SVP always have lower energy than SIVC, likely due to the winding energy penalty of SIVC hybridizing two Chern bands with opposite Chern number \cite{bultinck2020mechanism}. The energy splitting, however, is comparable to the energy splitting between the metallic state and the AHC state.

Finally, we make a few comments based on the SCHF data.
In RMG, the energy splitting due to different mechanisms form the following hierarchy, 
\begin{equation}
    \Delta_{\mathrm{Stoner}} \gg \Delta_{\mathrm{AHC}} \approx \Delta_{\mathrm{IVC}} \gg \Delta_{\text{moir\'{e}}}
\end{equation}
The system first forms a flavor order due to Stoner ferromagnetism, and then interaction drives the `quarter metal' to an insulator with finite Chern number, while disfavoring the IVC state by $\Delta_{\mathrm{IVC}}$. The moir\'{e} potential is not strong enough to alter the energy competition at $\nu=1$.


\subsection{Identification of moir\'{e}-proximate side from $\nu=4$ SCHF}
\label{app:nu_4_identification}

To confirm whether our setup of moir\'{e}-proximate/distant side corresponds to that in experiments, we identify the experiment observation of correlated insulator at $N_L=5,\nu=4$~\cite{lu2023fractional}: At low displacement field $|D|<\SI{0.5}{V}\mathrm{/nm}$, corresponding to $u_D\sim \SI{30}{meV}$, an insulator is found at $D<0$ side up to $|D|=\SI{0.4}{V}\mathrm{/nm}$, but not at $D>0$. We now will show that this agrees with our numerical observation at similar $u_D$, establishing a matching between the two works.

We perform self-consistent Hartree-Fock (SCHF) with both spin and valley degrees of freedom. We only include the conduction band as in the main text, and restrict ourselves to $u_D\ge\SI{15}{meV}$ where the valence bands are sufficiently separated. At $\nu=4$, we search for the ground state in the manifold without intervalley coupling, which excludes intervalley coherent (IVC) orders.

\begin{figure}[h]
	\centering
	\includegraphics[width=0.35\textwidth]
    {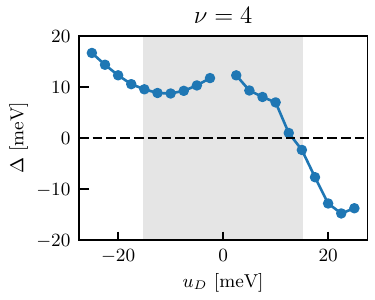}
    \caption{Charge gap of SCHF ground state at $\nu=4$ for $N_L=5$. Model parameters follow Appendix~\ref{app:numerical_setup}, but include both spins and valleys. $|u_D|<15\,\mathrm{meV}$ results are shaded since they are less reliable in SCHF. Parameters: $\theta=0.77^\circ$, $30\times 30$ unit cells.}
    \label{fig:app_gap_uD}
\end{figure}

We examine the resulting charge gap $\Delta$ at different $u_D$, as shown in Fig.~\ref{fig:app_gap_uD}, which shows a clear asymmetry between $u_D>0$ and $u_D<0$: a charge gap is only identified for $u_D<0$ but not for $u_D>0$, which agrees with the experiment described above. This suggests that the experiment identifies $\nu=4$ insulator only on the moir\'{e}-proximate regime. Although there could be other competing orders unconsidered here, the identification of the gapless phase is expected to be robust even beyond Hartree-Fock approximations.

\subsection{Details of $N_L$-$u_D$ phase diagram}
\begin{figure}[h]
	\centering
	\includegraphics[width=0.48\textwidth]{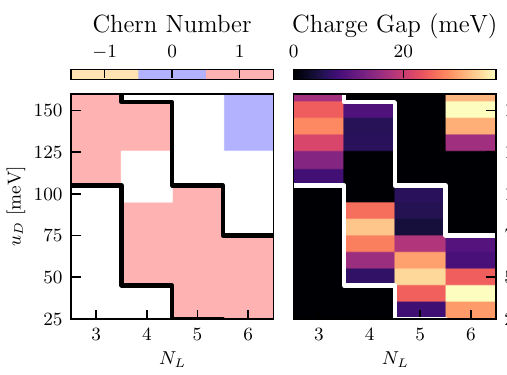}
    \caption{(a) Chern-number and (b) charge-gap phase diagram at $\nu=1$ for rhombohedral graphene with $N_L = 3-6$ and $u_D=0-160\,\mathrm{meV}$, with fixed $\theta=0.77^\circ$ and $\epsilon_r=5$ at each $N_L$. The boundary of $C=-1$ AHC phase is highlighted. The white region in the Chern-number plot denotes metal with a direct or indirect band gap. Parameters: $\theta=0.77^\circ$, $36\times 36$ unit cells.}
    \label{fig:app_Nlayer_uD}
\end{figure}

Fig.~\ref{fig:app_Nlayer_uD} shows the extended $N_L$-$u_D$ phase diagram in Fig. \ref{fig:intro}(c) which we extend up to $u_D=\SI{160}{meV}$. Each $N_L$ shows similar patterns as $N_L=5$ discussed in Fig. \ref{fig:phase_details}(a). First, a broad regime of $C=-1$ AHC phase is stabilized at moderate $u_D$.
The lower critical $u_D$, referred to as $u_D^*$ in the main text, decreases with $N_L$, indicating the stronger relevance of AHC in higher-$N_L$ rhombohedral graphenes. Meanwhile, the higher critical $u_D$ is controlled by the development of an indirect band gap between $C=-1$ band and higher bands, which leads to a first-order transition into metal (and occasionally the ``reentrance" of $C=-1$ state at $N_L=4$.) At even higher $u_D$, other phases such as the $C=0$ insulator take over, but given the stronger band mixing it is necessary to include more bands in SCHF to accurately resolve the phase competition. We therefore caution against trusting the phase diagram at large $u_D$.

\begin{figure}[h!]
	\centering
	\includegraphics[width=0.4\textwidth]{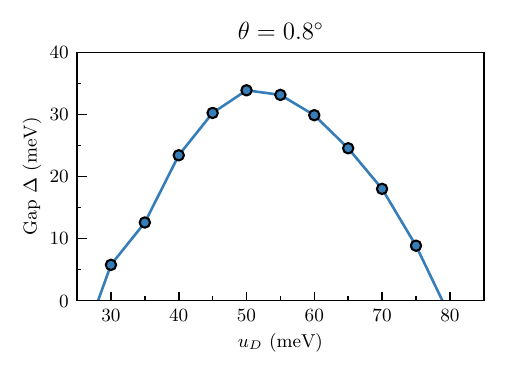}
    \caption{Dependence of charge gap on $u_D$ at $N_L=5$, $\theta=0.8^\circ$ and $\epsilon_r=5$. We take a $30\times 30$ unit cell. A clear non-monotonic dependence of the charge gap on $u_D$ is observed.}
    \label{fig:app_gap_uD_nu_1}
\end{figure}



We note that the charge gap, shown in Figs.~\ref{fig:app_Nlayer_uD} and~\ref{fig:app_gap_uD_nu_1}, has sensitive and non-monotonic dependence on displacement $u_D$. However, it is not directly proportional to the moir\'e potential, shown in Fig.~\ref{fig:robustness}(a) of the main text. Here we also comment on the observable effects of the charge gap. In addition to the thermal activation behavior expected in transport measurements, the width of the plateau with quantized Hall conductance also depends on the charge gap~\cite{LaughlinDisorder, DavidNote}.. If we assume that impurities induce subgap localized states with a constant density of states, the Hall conductance will remain quantized until the doped charge fully fills all subgap states, whose total density is proportional to the charge gap. In Ref.~\cite{lu2023fractional}, the plateau width of the $\nu=1$ Chern insulator also behaves non-monotonically with the displacement field, possibly related to this behavior of the charge gap.

\subsection{Effect of Graphene-Scale Parameters}

This section considers the effect of the graphene-scale parameters on our results. An important phenomenon in graphene and its related systems is the renormalization of the Fermi velocity by Coulomb interactions, which has been observed experimentally~\cite{castro2009the,hwang2011direct,yang2021experimental}. Namely, the effective velocity (i.e. $t_0$) increases as the energy scale or filling approaches charge neutrality. When discussing phenomena such as those here that depend on both a small moir\'e scale and a much larger displacement field scale, it is unclear which Fermi velocity to employ. Estimates in the literature for $u_D = 0$ vary between $t_0 \approx \SI{2.6}{eV}$ and $t_0 \approx \SI{3.1}{eV}$, depending on the relevant energy scales, number of layers, and other details~\cite{zhou2021half, zhang2010band}. Moreover, this problem is exacerbated by the fact that one must use a ``bare" Fermi velocity in models to which interactions are added, but renormalization of the velocity is a beyond-mean-field effect. 

\begin{figure}
    \centering
    \includegraphics{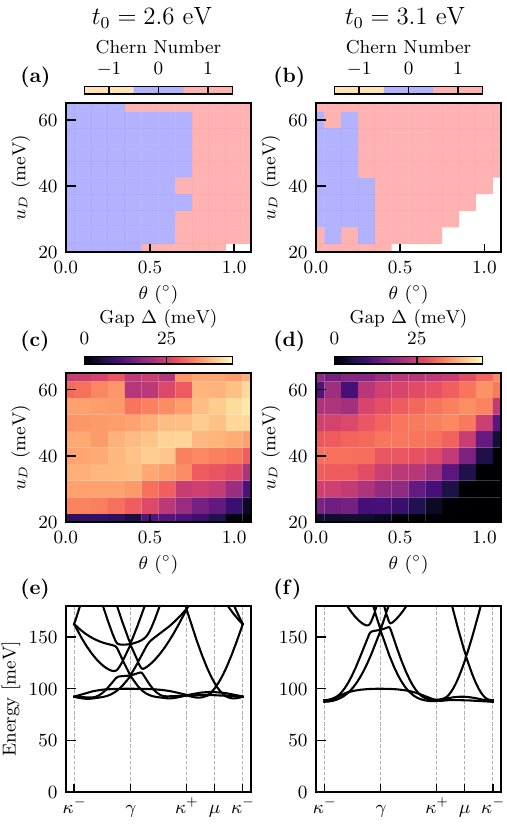}
    \caption{The effect of large versus small Fermi velocity is minimal. (a,b) SCHF phase diagrams at $\nu=1$ as a function of twist angle $\theta$ and vertical displacement field $u_D$, showing the difference between large Fermi velocity, corresponding to $t_0 = \SI{3100}{meV}$, versus small Fermi velocity with $t_0 =\SI{2600}{meV}$. Here colors show the Chern number of insulators, and gray data points have gaps of \SI{2}{meV} or less. (c,d) Charge gaps above $\nu=1$ within SCHF. (e,f) Single-particle ($\nu=0$) bandstructures at large and small Fermi velocity. Note that the gap to the second band at $\gamma$ increases dramatically at large $t_0$.
    }
    \label{fig_app:vary_t0_SCHF_phase_diagrams}
\end{figure}

We now show changing the Fermi velocity dramatically shifts the phase boundaries but not the phases present. Fig.~\ref{fig_app:vary_t0_SCHF_phase_diagrams} shows phase diagrams at $\nu=1$ within SCHF for $t_0 = \SI{2.6}{eV}$ and $\SI{3.1}{eV}$, respectively. The latter option is used in the main text and throughout the rest of this work. Given that the various ground states candidates are separated on the $O(1)$\si{meV} scale, this \SI{500}{meV} perturbation has a remarkably small effect. Comparing Fig.~\ref{fig_app:vary_t0_SCHF_phase_diagrams} (a) and (b), one sees that the smaller $t_0$ favors $C=0$ over $C=1$.
This remarkable resilience of the Chern band to these parameter changes suggests a much smaller number of hidden parameters that control the phase diagram. Future work will explicate this phenomenon.

\begin{figure}[h]
	\centering
	\includegraphics[width=0.35\textwidth]
    {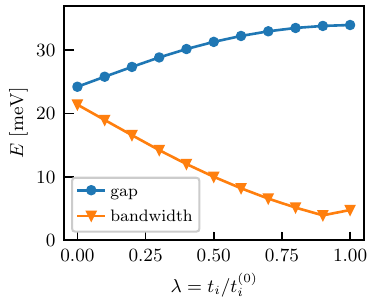}
    \caption{Charge gap and bandwidth of RMG ground states with modified hopping strength $\lambda t_{2,3,4}, \lambda\in(0,1)$. All ground states are found to be AHC phase with $C=-1$. Parameters: $(u_D, \theta)=(\SI{50}{meV}, 0.77^\circ)$, $30\times 30$ unit cells.}
    \label{fig:app_t_scan}
\end{figure}

In addition, it turns out that even large differences in the graphene-scale parameters do not strongly affect our results at $\nu=1$. Fig.~\ref{fig:app_t_scan} demonstrates the robustness of AHC as the higher hopping terms are varied. Explicitly, we replace hopping terms $t_{2,3,4}$ in Eq.\,\eqref{eq:rhombohedral_graphene} with $\lambda t_{2,3,4}$, where $\lambda=1$ correspond to realistic RMG and $\lambda=0$ correspond to simplified RMG with only nearest-neighbor hoppings. We then compute its ground state at $\lambda\in(0,1)$, which all result in AHC phases with $C=-1$. The charge gap of the ground state remains large through the tuning, suggesting the stability of the phase. The bandwidth of the $C=-1$ band gradually increases at lower $\lambda$, which indicates its deviation from the optimal condition realizing FQAH.

\section{Details of the Many-Body Numerics at Fractional Filling}

To determine the many-body ground state at fractional filling, we undertake a two-step procedure: (1) solve for the SCHF ground state at filling $\nu=1$ and then (2) solve the many-body problem restricted to the lowest SCHF band (which is spin and valley-polarized) at fractional filling using either exact-diagonalization or the density-matrix renormalization group numerics.

An explicit assumption of such numerics is spin- and valley-polarization within the whole range $0 \le \nu \le 1$.
When should this procedure be valid? It is first helpful to review the experimental phenomena. Ref~\cite{lu2023fractional} reported Laughlin and Jain sequence states at fillings $\nu=2/5,3/7,4/9,4/7,3/5$ as well as $\nu=2/3$. Crucially, they also reported that the region of large Hall conductance only extends down to $\nu=0.2-0.4$ (depending on $u_D$), whereupon it transitions to a ``speckled" region that may be due to Wigner crystallization. Our assumption that the many-body ground state is within the many-body Hilbert space of the $\nu=1$ Chern band is clearly violated at this point, rendering our model inapplicable for sufficiently low fillings. 
It is also helpful to note the simpler case of the lowest Landau level, where the fractional state at $\nu=2/3$ can be spin unpolarized~\cite{PhysRevB.41.7910,PhysRevLett.72.3405,PhysRevB.53.15845} despite a spin-polarized LL being favored at $\nu=1$.

It is unclear how to perform the corresponding computation without spin polarization, as it is unclear what the 2-band low-energy (i.e. 1 band with spin) Hilbert space should be considered. One \textit{cannot} simply add spin to the $\nu=1$ Chern band; instead one must restrict to a spinful Hilbert space from the full Hilbert space. Determining such a minimal model may require a four-flavor many-body calculation with many-band and is beyond the scope of the current work. Nevertheless, we expect that spin- and valley-polarized states are at least low-energy candidate ground states through most of the range. We therefore work within the one-band model to examine the presence of Jain sequence states in the simplest scenario.

Explicitly, this is done as follows. Consider the SCHF Hamiltonian $h_{HF}[P]$ at $\nu=1$, which may be diagonalized as
\begin{equation}
    h_{HF}[P]_{\alpha \beta}(\v{k}) U_{\beta \gamma}(\v{k}) = U_{\alpha \gamma}(\v{k}) E_{\gamma}(\v{k}).
\end{equation}
We may then pass to the basis of Hartree-Fock bands
\begin{equation}
 \ket{\phi_{\v{k} \alpha}} = \ket{\psi_{\v{k}\beta}} U_{\beta \alpha}(\v{k}),
\end{equation}
which constitutes a unitary rotation of our set of $N_b$ low-energy bands. We may then integrate out the $N_b-1$ bands above the $\nu=1$ gap at mean-field level --- which again is simply an index restriction --- to arrive at a one-band Hamiltonian
\begin{equation}
    \hat{H} = \sum_{\v{k}} t_{\v{k}} \hat{n}_{\v{k}} + \frac{1}{2A} \sum_{\v{q}, \v{k},\v{k}'} U_{\v{q}} \lambda_{\v{q}}(\v{k}) \lambda_{-\v{q}}(\v{k}') \hat{d}_{\v{k}}^\dagger \hat{d}_{\v{k}'}^\dagger \hat{d}_{\v{k}'-\v{q}} \hat{d}_{\v{k}+\v{q}},
    \label{eq:app_1_band_Hamiltonian}
\end{equation}
where $\hat{d}_{\v{k}} \ket{0} =  \ket{\phi_{\v{k} 0}}$ is a creation operator for the lowest SCHF band, a band with form factors $\lambda_{\v{q}}(\v{k}) = \braket{\phi_{\v{k}0}|e^{-i\v{q} \cdot \v{r}}| \phi_{\v{k}+\v{q} 0}}$, and $\hat{n}_{\v{k}} =  \hat{d}_{\v{k}}^\dagger \hat{d}_{\v{k}}$. Note that while we have projected into the Hilbert space of the first SCHF band, the kinetic energy of the band is determined relative to $\hat{H}_{\mathrm{RMG}}$ at $\nu=0$ to avoid double-counting interactions.

\subsection{Exact Diagonalization at Fractional Filling}

\begin{figure*}
    \centering
    \includegraphics{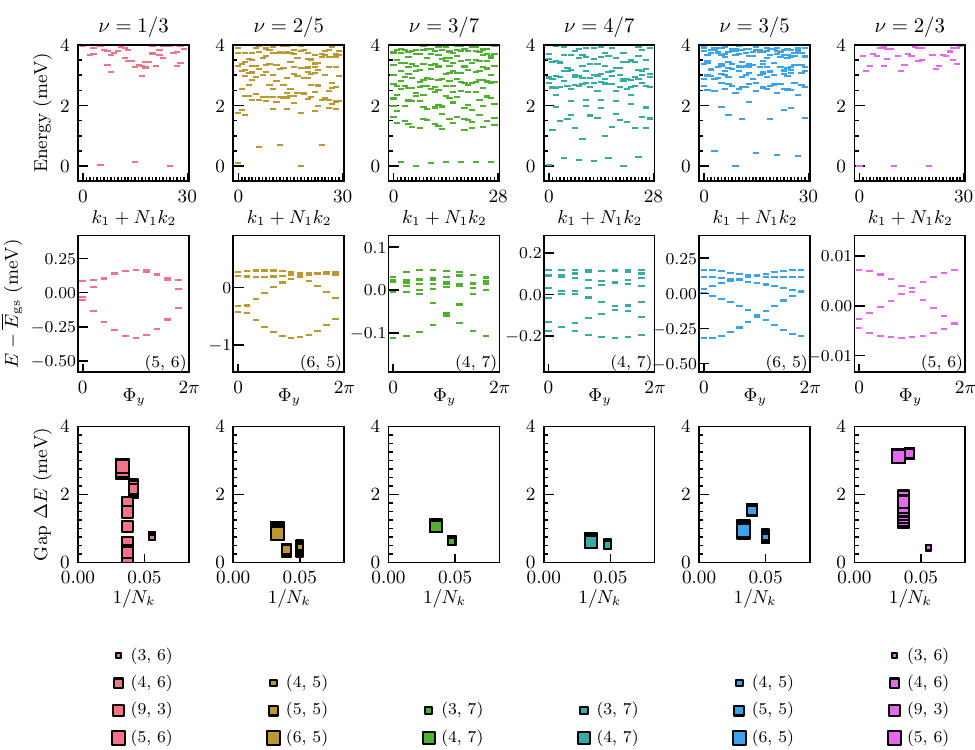}
    \caption{Exact diagonalization at various fractional fillings over the lowest $\nu=1$ band of RMG. Top row: spectra at the largest system sizes available at each filling. Second row: spectral flow under flux threading. Third row: system size dependence, showing the gap between the quasidegenerate ground states and the first excited state for all fluxes at each available system size. Details in text. Parameters: $\theta=0.77^\circ$, $u_D =\SI{50}{meV}$, other parameters match Fig. 2 of the main text}
    \label{fig_app:ED_finite_size}
\end{figure*}

To perform exact diagonalization, we compute Eq. \eqref{eq:app_1_band_Hamiltonian} on a momentum space Monkhorst-Pack grid $\v{k}[n_1,n_2]$ of size $N_{k_1} \times N_{k_2}$. We then subsample that grid on points
\begin{equation}
    \setc{\v{k}[s_1 m_1 + \phi_1, s_2 m_2]}{0 \le m_1 < M_1, 0 \le m_2 < M_2},
\end{equation}
to produce a grid of size $M_{1} \times M_{2}$ where the integer ratios $s_i$ obey $s_1 M_1 = N_{k_1}$ and $s_{2} M_2 = N_{k_2}$. Here $0 \le \phi_1 < s_1$ is an offset in the subsampling to enable flux threading. For the main text, we take $N_{k_1} \times N_{k_2} = 36 \times 36$ and $M_1 \times M_2 = 4 \times 6$.

Exact diagonalization was performed at a range of fractional fillings to determine their suitability for fractional quantum Hall states: $\nu=1/3,2/5,3/5,4/7,3/5,2/3$. At each filling we computed spectrum over a range of system sizes and performed flux threading along $k_x$ for the largest system size available, with $N_k = N_{k_1} \times N_{k_2}$ up to $30$. The results are shown in Fig.~\ref{fig_app:ED_finite_size}.

The top row of Fig.~\ref{fig_app:ED_finite_size} shows the low-lying spectrum at all center of mass momenta for each system size. In each case, we observe a quasi-degenerate ground states with degeneracy $q$ at filling $\nu=p/q$, with a gap of \SIrange{1}{3}{meV}, depending on filling. This suggests that the ground states here are all fractional Chern insulators of Laughlin and Jain type. 

To support this inference, the second row of Fig.~\ref{fig_app:ED_finite_size} shows the spectral flow under flux threading for each filling. To make the permutation of states clear, we plot energies $E - \overline{E}_{\mathrm{gs}}$, where $\overline{E}_{\mathrm{gs}}(\Phi_y)$ is the average energy of the quasidegerate ground states at each flux $\Phi_y$. For instance $\nu=2/3$, the spectral flow clearly permutes the ground states on a $5 \times 6$ site cluster (see main text on $4\times 6$ sites). Indeed, we see a permutation of the quasidegenerate ground states for each filling for at least some system sizes. Explicitly, we see such a spectral flow for $5 \times 6$ at $\nu=1/3$, but the $9\times 3$ cluster is gapless at the same filling. This flux pumping evidence supports the inference that FCIs of Laughlin and Jain types are ground state competitors at each filling we consider. However, they may not be the true ground state at all fillings in the thermodynamic limits.

Finally, the bottom row of Fig.~\ref{fig_app:ED_finite_size} shows the system size dependence of the gaps. For each system size, we show the gap between the top quasidegenerate ground state and the next state in the spectrum at the same filling with any momentum. Each box represents that gap at one value of the flux, and the sizes of the boxes increase with system size. We note that the system size dependence is highly non-monotonic and far from the scaling regime, suggesting that finite size effects dominate the results --- as expected for a system with such concentrated Berry curvature pockets and other quantum geometric features. Of particular note is the system sizes $9\times 3$ at $\nu=1/3$ filling, where we observe gap closings under spectral flow. This suggests a qualitatively different state is stabilize at these oblong aspect ratios, whose energy per electron is slightly less than the competing FCI state (at these small system sizes). 

Overall we may conclude that FCIs are low-energy ground state \textit{candidates} at all fillings $\nu=1/3,2/5,3/7,4/7,3/5,2/3$. However, the many-body gaps for these states are small at $\nu=2/5$, and can vanish entirely at $\nu=1/3$, depending on the cluster considered. Moreover, the gaps have not stabilized on even the largest sizes we could consider, and we find small changes in parameters such as $u_D$ often favor CDW orders or Fermi liquids. Finally, modelling uncertainty greatly exceeds finite-size effects here. The balance of the evidence supports the following conclusions: filling $1/3$ is inconclusive; $2/5$ and $4/7$ are likely FCIs with small gaps, and $3/7$,$3/5$, $2/3$ have FCI ground states within our model.

\subsection{DMRG at Fractional Filling}

\begin{figure}
    \centering
    \includegraphics{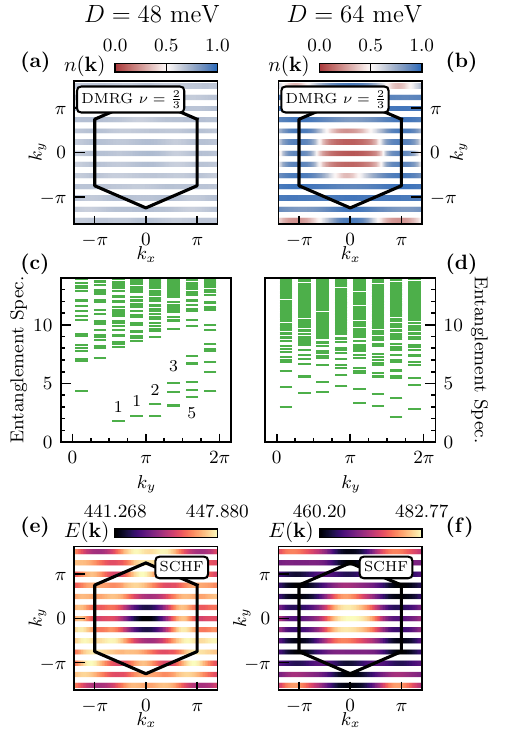}
    \caption{DMRG results at $u_D = \SI{48}{meV}$ (left) and \SI{64}{meV} (right). Here (a,b) give occupations over the Brillouin zone, (c,d) give entanglement spectra, and (e,f) show the dispersion of the corresponding lowest SCHF Chern band at $\nu=1$. See text for details.}
    \label{fig_app:DMRG}
\end{figure}

To apply the density-matrix renormalization group, we follow previous work on moir\'e DMRG systems ~\cite{MPOCompression2, parker2021strain, Dan_parker21, wang2022kekul} to place the system onto an infinite cylinder geometry. We start with a rectangular moir\'e Brillouin zone, shown in Fig. \ref{fig:mbz_geometry}, with side lengths $B_1 = \n{\v{g}_1}, B_2 = \frac{\sqrt{3}}{2} \n{\v{g}_2}$. We impose periodic boundary conditions in the ``2" direction corresponding to momentum discretization
\begin{equation}
  k_2[n] = B_2 \times \begin{cases}
    -\frac{1}{2} + \frac{n + \frac{\Phi_2}{2\pi}}{N_2} & \text{if } N_2 \equiv 0 \pmod 2\\[0.5em]
    -\frac{1}{2} + \frac{n+ \frac{\Phi_2}{2\pi}+\frac{1}{2}}{N_2} & \text{if }  N_2 \equiv 1 \pmod 2 \\
  \end{cases},
 \label{eq:DMRG_wire_momenta}
\end{equation}
relative to an origin at moir\'e-$\gamma$. This gives $N_2$ evenly spaced ``wires" across the Brillouin zone, which intersect the $\gamma$ point for flux $\Phi_2=0$.

We then take a basis of hybrid Wannier functions along each wire, i.e. maximally localized along $\v{a}_1$ and periodic along $\v{a}_2$:
\begin{equation}
    \ket{w_{n,k_2}} = \frac{1}{\sqrt{N_1}} \sum_{\v{k}_1} \ket{\varphi_{\v{k}}} e^{-i n \v{k}\cdot\v{a}_1}; \quad \ket{\varphi_{\v{k}}} = \ket{\phi_{\v{k} 0}} e^{i \zeta(\v{k})}.
\end{equation}
Here the gauge choice $\zeta(\v{k})$ is chosen such that
\begin{subequations}
    \begin{align}
        \hat{T}_{1} \ket{w^{n,k_2}} &= \ket{w^{n-1,k_2}}\\
        \hat{T}_{2} \ket{w^{n,k_2}} &= e^{i 2 \pi k_2/B_2} \ket{w^{n,k_2}}\\
        \hat{P} e^{-i \hat{r}_1 B_1} \hat{P} \ket{w^{n,k_2}} &= e^{i 2\pi P_1(k_2)} \ket{w^{n,k_2}}
    \end{align}
\end{subequations}
where the polarizations $P_1(k_2) = \frac{B_1}{2\pi} \braket{w^{0,k_2}|\hat{r}_1|w^{0,k_2}}$ are the centers of the Wannier orbitals in the first unit cell, in accordance with the modern theory of polarization~\cite{vanderbilt_2018}. The Hamiltonian \eqref{eq:app_1_band_Hamiltonian} can be straightforwardly expressed in terms of fermion operators $\hat{f}_{n,k_2}^\dagger \ket{0} = \ket{w_{n,k_2}}$ on an infinite cylinder cylinder geometry. As this is a long-ranged, two-dimensional Hamiltonian, working with it efficiently as a matrix product operator (MPO) requires the technique of ``MPO Compression'' developed by some of us ~\cite{MPOCompression1,MPOCompression2}, which creates a faithful low-rank approximation of the Hamiltonian with errors below the $10^{-2}$ \si{meV} level.

We then apply standard DMRG using the open-source tenpy~\cite{tenpy} library. We work at cylinder circumference $L_y=8$. For FCI states, we converge the ground state to norm error below $10^{-13}$ on bond dimensions up to $\chi = 1536$  --- since the state is fully gapped, only moderate bond dimensions are required  To diagnose the fractional Chern insulator we use the entanglement spectrum technique first proposed by~\cite{li2004entanglement}, finding the characteristic degeneracy counting matching partition numbers $1,1,2,3,5,7,...$ for each bond. We have also verified this holds for $L_y=5, 6$.

Fig. \ref{fig_app:DMRG} compares the results of DMRG at two difference displacement fields, $u_D = \SI{48}{meV}$ and $u_D = \SI{64}{meV}$. The first is selected as the location where the FCI gap is maximized in ED. The second is in a regime where the $\nu=1$ state is still a Chern band, but whose bandwidth is now $\sim{} \SI{20}{meV}$ and trace condition violation is $\approx 1$. Both systems are at $\theta=0.77^\circ$ and all other parameters are identical. 

Fig.~\ref{fig_app:DMRG}(a,b) shows the occputaion numbers $n(\v{k})~=~\braket{\hat{c}^\dagger_{\v{k}} \hat{c}_{\v{k}}}$ across the Brillouin zone for both displacement fields. For $u_D = \SI{48}{meV}$, $n(\v{k}) = \frac{2}{3} \pm 0.04$ over the whole Brillouin zone, whereas for $u_D = \SI{64}{meV}$ the occcupations $n(\v{k})$ vary from $0.08$ to $0.95$ at $\chi = 1536$. Moreover, the occupations are almost entirely depleted near $\gamma$, precisely where the SCHF bandstructure, Fig.~\ref{fig_app:DMRG}(f), is peaked. We further check that the charge $0$ and charge $1$ correlation lengths are increasing with bond dimension, which gives further confirmation of gapless modes in the these channels as expected from a Fermi liquid. The entanglement spectrum Fig.~\ref{fig_app:DMRG}(d) does not have a clear low-lying branch separated from the rest of the spectrum, and is clearly not chiral with the expected CFT degeneracies. The large displacement field system at $u_D = \SI{64}{meV}$ is therefore a Fermi liquid. However, we caution that this Fermi liquid from doping the $\nu=1$ Chern band may not correspond to the experimentally-observed Fermi liquid at larger displacement fields.

\end{document}